\documentclass[onecolumn]{aastex631}

\newcommand{\des}{DES}
\newcommand{\hst}{HST}
\newcommand{\wfc}{WFC3}
\newcommand{\gaia}{\textit{Gaia}}

\usepackage{natbib} 
\usepackage{amsmath}
\usepackage{enumitem}
\usepackage{verbatim}
\usepackage{graphicx}
\usepackage{subfigure}
\usepackage{color}
\usepackage{xcolor}
\usepackage{float}
\usepackage{footnote}
\usepackage{hanging}
\usepackage{hyperref}

\shorttitle{Synchronous rotation of Eris}

\reportnum{DES-2022-0747}
\reportnum{ FERMILAB-PUB-23-122-PPD}

\begin{document}

\title{Synchronous rotation in the (136199) Eris-Dysnomia system}

\author[0000-0002-8613-8259]{Gary M. Bernstein}
\affiliation{Department of Physics and Astronomy, University of Pennsylvania, Philadelphia, PA 19104, USA}
\email{garyb@physics.upenn.edu}
\correspondingauthor{Gary M. Bernstein}

\author[0000-0002-6117-0164]{Bryan J. Holler}
\affiliation{Space Telescope Science Institute, Baltimore, MD 21218, USA}

\author[0009-0009-2907-5421]{Rosario Navarro-Escamilla}
\affiliation{Department of Physics and Astronomy, University of Pennsylvania, Philadelphia, PA 19104, USA}

\author[0000-0003-0743-9422]{Pedro H. Bernardinelli}
\altaffiliation{DiRAC Postdoctoral Fellow}
\affiliation{DIRAC Institute and the Department of Astronomy, University of Washington, 3910 15th Avenue NE, Seattle, WA 98195, USA}


\author{T.~M.~C.~Abbott}
\affiliation{Cerro Tololo Inter-American Observatory, NSF's National Optical-Infrared Astronomy Research Laboratory, Casilla 603, La Serena, Chile}
\author{M.~Aguena}
\affiliation{Laborat\'orio Interinstitucional de e-Astronomia - LIneA, Rua Gal. Jos\'e Cristino 77, Rio de Janeiro, RJ - 20921-400, Brazil}
\author{S.~Allam}
\altaffiliation{Deceased}
\affiliation{Fermi National Accelerator Laboratory, P. O. Box 500, Batavia, IL 60510, USA}
\author{O.~Alves}
\affiliation{Department of Physics, University of Michigan, Ann Arbor, MI 48109, USA}
\author{F.~Andrade-Oliveira}
\affiliation{Department of Physics, University of Michigan, Ann Arbor, MI 48109, USA}
\author{J.~Annis}
\affiliation{Fermi National Accelerator Laboratory, P. O. Box 500, Batavia, IL 60510, USA}
\author{D.~Bacon}
\affiliation{Institute of Cosmology and Gravitation, University of Portsmouth, Portsmouth, PO1 3FX, UK}
\author{D.~Brooks}
\affiliation{Department of Physics \& Astronomy, University College London, Gower Street, London, WC1E 6BT, UK}
\author{D.~L.~Burke}
\affiliation{Kavli Institute for Particle Astrophysics \& Cosmology, P. O. Box 2450, Stanford University, Stanford, CA 94305, USA}
\affiliation{SLAC National Accelerator Laboratory, Menlo Park, CA 94025, USA}
\author{A.~Carnero~Rosell}
\affiliation{Instituto de Astrofisica de Canarias, E-38205 La Laguna, Tenerife, Spain}
\affiliation{Laborat\'orio Interinstitucional de e-Astronomia - LIneA, Rua Gal. Jos\'e Cristino 77, Rio de Janeiro, RJ - 20921-400, Brazil}
\affiliation{Universidad de La Laguna, Dpto. Astrofísica, E-38206 La Laguna, Tenerife, Spain}
\author{J.~Carretero}
\affiliation{Institut de F\'{\i}sica d'Altes Energies (IFAE), The Barcelona Institute of Science and Technology, Campus UAB, 08193 Bellaterra (Barcelona) Spain}
\author{L.~N.~da Costa}
\affiliation{Laborat\'orio Interinstitucional de e-Astronomia - LIneA, Rua Gal. Jos\'e Cristino 77, Rio de Janeiro, RJ - 20921-400, Brazil}
\author{M.~E.~S.~Pereira}
\affiliation{Hamburger Sternwarte, Universit\"{a}t Hamburg, Gojenbergsweg 112, 21029 Hamburg, Germany}
\author{J.~De~Vicente}
\affiliation{Centro de Investigaciones Energ\'eticas, Medioambientales y Tecnol\'ogicas (CIEMAT), Madrid, Spain}
\author{S.~Desai}
\affiliation{Department of Physics, IIT Hyderabad, Kandi, Telangana 502285, India}
\author{P.~Doel}
\affiliation{Department of Physics \& Astronomy, University College London, Gower Street, London, WC1E 6BT, UK}
\author{A.~Drlica-Wagner}
\affiliation{Department of Astronomy and Astrophysics, University of Chicago, Chicago, IL 60637, USA}
\affiliation{Fermi National Accelerator Laboratory, P. O. Box 500, Batavia, IL 60510, USA}
\affiliation{Kavli Institute for Cosmological Physics, University of Chicago, Chicago, IL 60637, USA}
\author{S.~Everett}
\affiliation{Jet Propulsion Laboratory, California Institute of Technology, 4800 Oak Grove Dr., Pasadena, CA 91109, USA}
\author{I.~Ferrero}
\affiliation{Institute of Theoretical Astrophysics, University of Oslo. P.O. Box 1029 Blindern, NO-0315 Oslo, Norway}
\author{J.~Frieman}
\affiliation{Fermi National Accelerator Laboratory, P. O. Box 500, Batavia, IL 60510, USA}
\affiliation{Kavli Institute for Cosmological Physics, University of Chicago, Chicago, IL 60637, USA}
\author{J.~Garc\'ia-Bellido}
\affiliation{Instituto de Fisica Teorica UAM/CSIC, Universidad Autonoma de Madrid, 28049 Madrid, Spain}
\author{D.~W.~Gerdes}
\affiliation{Department of Astronomy, University of Michigan, Ann Arbor, MI 48109, USA}
\affiliation{Department of Physics, University of Michigan, Ann Arbor, MI 48109, USA}
\author{D.~Gruen}
\affiliation{University Observatory, Faculty of Physics, Ludwig-Maximilians-Universit\"at, Scheinerstr. 1, 81679 Munich, Germany}
\author{G.~Gutierrez}
\affiliation{Fermi National Accelerator Laboratory, P. O. Box 500, Batavia, IL 60510, USA}
\author{K.~Herner}
\affiliation{Fermi National Accelerator Laboratory, P. O. Box 500, Batavia, IL 60510, USA}
\author{S.~R.~Hinton}
\affiliation{School of Mathematics and Physics, University of Queensland,  Brisbane, QLD 4072, Australia}
\author{D.~L.~Hollowood}
\affiliation{Santa Cruz Institute for Particle Physics, Santa Cruz, CA 95064, USA}
\author{K.~Honscheid}
\affiliation{Center for Cosmology and Astro-Particle Physics, The Ohio State University, Columbus, OH 43210, USA}
\affiliation{Department of Physics, The Ohio State University, Columbus, OH 43210, USA}
\author{D.~J.~James}
\affiliation{Center for Astrophysics $\vert$ Harvard \& Smithsonian, 60 Garden Street, Cambridge, MA 02138, USA}
\author{K.~Kuehn}
\affiliation{Australian Astronomical Optics, Macquarie University, North Ryde, NSW 2113, Australia}
\affiliation{Lowell Observatory, 1400 Mars Hill Rd, Flagstaff, AZ 86001, USA}
\author{N.~Kuropatkin}
\affiliation{Fermi National Accelerator Laboratory, P. O. Box 500, Batavia, IL 60510, USA}
\author{J.~L.~Marshall}
\affiliation{George P. and Cynthia Woods Mitchell Institute for Fundamental Physics and Astronomy, and Department of Physics and Astronomy, Texas A\&M University, College Station, TX 77843,  USA}
\author{J. Mena-Fern{\'a}ndez}
\affiliation{Centro de Investigaciones Energ\'eticas, Medioambientales y Tecnol\'ogicas (CIEMAT), Madrid, Spain}
\author{R.~Miquel}
\affiliation{Instituci\'o Catalana de Recerca i Estudis Avan\c{c}ats, E-08010 Barcelona, Spain}
\affiliation{Institut de F\'{\i}sica d'Altes Energies (IFAE), The Barcelona Institute of Science and Technology, Campus UAB, 08193 Bellaterra (Barcelona) Spain}
\author{R.~L.~C.~Ogando}
\affiliation{Observat\'orio Nacional, Rua Gal. Jos\'e Cristino 77, Rio de Janeiro, RJ - 20921-400, Brazil}
\author{A.~Pieres}
\affiliation{Laborat\'orio Interinstitucional de e-Astronomia - LIneA, Rua Gal. Jos\'e Cristino 77, Rio de Janeiro, RJ - 20921-400, Brazil}
\affiliation{Observat\'orio Nacional, Rua Gal. Jos\'e Cristino 77, Rio de Janeiro, RJ - 20921-400, Brazil}
\author{A.~A.~Plazas~Malag\'on}
\affiliation{Department of Astrophysical Sciences, Princeton University, Peyton Hall, Princeton, NJ 08544, USA}
\author{M.~Raveri}
\affiliation{Department of Physics, University of Genova and INFN, Via Dodecaneso 33, 16146, Genova, Italy}
\author{K.~Reil}
\affiliation{SLAC National Accelerator Laboratory, Menlo Park, CA 94025, USA}
\author{E.~Sanchez}
\affiliation{Centro de Investigaciones Energ\'eticas, Medioambientales y Tecnol\'ogicas (CIEMAT), Madrid, Spain}
\author{I.~Sevilla-Noarbe}
\affiliation{Centro de Investigaciones Energ\'eticas, Medioambientales y Tecnol\'ogicas (CIEMAT), Madrid, Spain}
\author{M.~Smith}
\affiliation{School of Physics and Astronomy, University of Southampton,  Southampton, SO17 1BJ, UK}
\author{M.~Soares-Santos}
\affiliation{Department of Physics, University of Michigan, Ann Arbor, MI 48109, USA}
\author{E.~Suchyta}
\affiliation{Computer Science and Mathematics Division, Oak Ridge National Laboratory, Oak Ridge, TN 37831}
\author{M.~E.~C.~Swanson}
\author{P.~Wiseman}
\affiliation{School of Physics and Astronomy, University of Southampton,  Southampton, SO17 1BJ, UK}

\collaboration{1000}{(The DES Collaboration)}
\suppressAffiliations

\makeatletter
\def\ps@pprintTitle{%
 \let\@oddhead\@empty
 \let\@evenhead\@empty
 \def\@oddfoot{\centerline{\thepage}}%
 \let\@evenfoot\@oddfoot}
\makeatother

\begin{abstract}
We combine photometry of Eris from a 6-month campaign on the Palomar 60-inch telescope in 2015, a 1-month Hubble Space Telescope \wfc\ campaign in 2018, and Dark Energy Survey data spanning 2013--2018 to determine a light curve of definitive period $15.771\pm 0.008$~days (1-$\sigma$ formal uncertainties), with nearly sinusoidal shape and peak-to-peak flux variation of 3\%.  This is consistent at part-per-thousand precision with the $P=15.78590\pm0.00005$~day period of Dysnomia's orbit around Eris, strengthening the recent detection of synchronous rotation of Eris by \citet{szakats} with independent data.  Photometry from \gaia\ are consistent with the same light curve.  We detect a slope of $0.05\pm0.01$~mag per degree of Eris' brightness with respect to illumination phase, intermediate between Pluto's and Charon's values.  Variations of $0.3$~mag are detected in Dysnomia's brightness, plausibly consistent with a double-peaked light curve at the synchronous period. The synchronous rotation of Eris is consistent with simple tidal models initiated with a giant-impact origin of the binary, but is difficult to reconcile with gravitational capture of Dysnomia by Eris.
\end{abstract}


\section{Introduction}
The trans-Neptunian region of icy minor bodies beyond the orbit of Neptune contains a record of the chemical composition and early dynamical history of the solar system. The observed dynamical structure of the modern trans-Neptunian region is complex \citep[e.g.,][]{Elliot2005, Gladman2008, Petit2011, Adams2014, Bannister2018,Bernardinelli2022} and appears to be the result of giant planet migration early in solar system history \citep[e.g.,][]{FI84,Gomes2005,HM2005,Morby2007,Walsh2011,Lawler2019}. Debate surrounds the timing and mechanism of this migration, but there is no debate that this period was a chaotic one for the primordial Kuiper Belt. A large fraction of the original mass in this disk was lost to the inner solar system, the Oort Cloud, or interstellar space \citep[e.g.,][]{Gomes2005,Dones2015,Malhotra2019}, and some of the surviving trans-Neptunian objects (TNOs) are members of binary or multiple systems \citep[e.g.,][]{Noll2020}. The existence of some of these binary and multiple systems could be due to interactions with objects perturbed onto crossing orbits by the migration of the giant planets.

\begin{figure}[ht!]
\begin{center}
\includegraphics[scale=0.64,trim=0cm 0cm 0cm 0cm,clip=true]{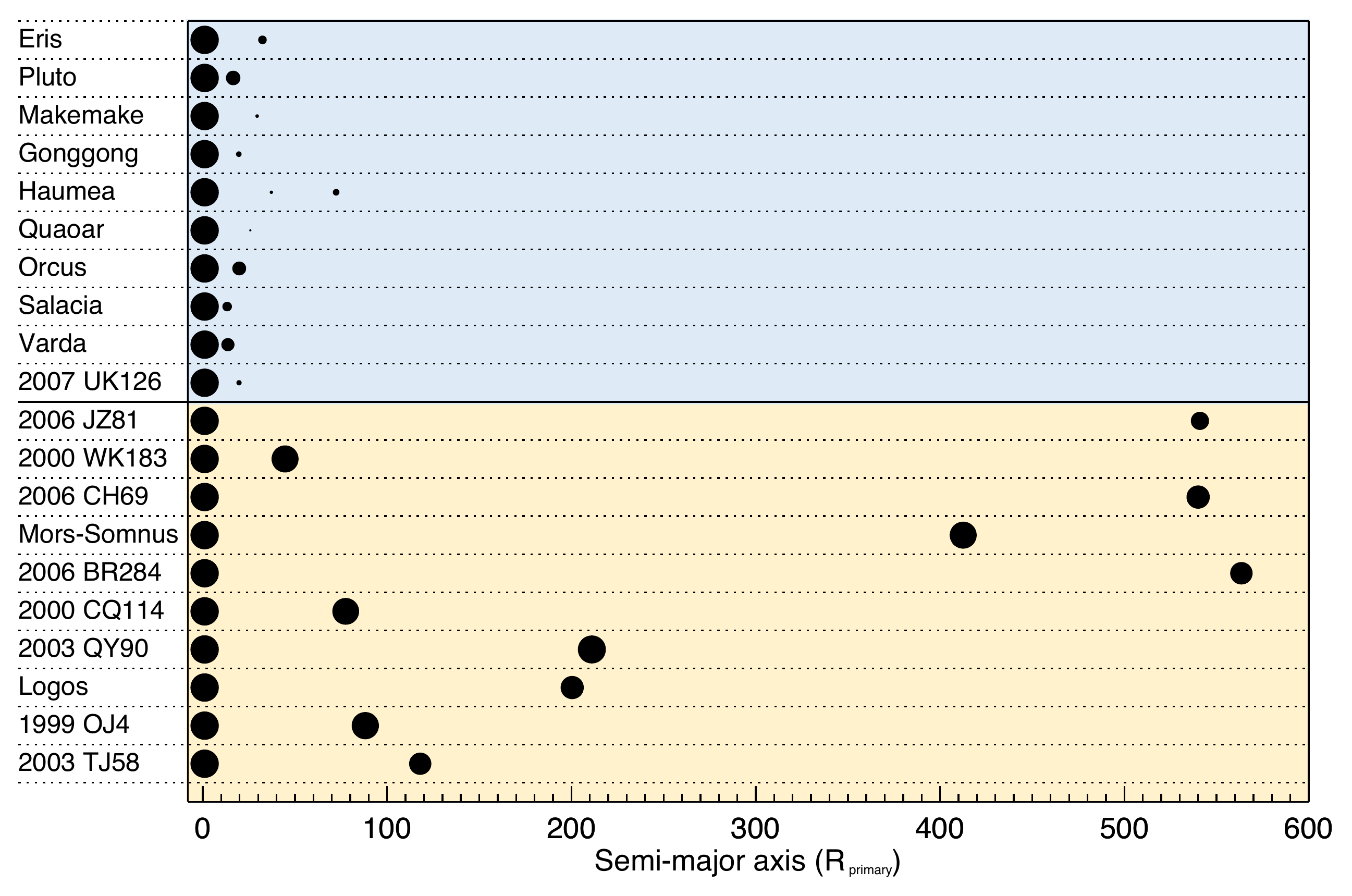}
\caption{Each row depicts the size and separation of a TNO satellite relative to its primary body.
Above the solid black line are shown 10 of the most massive known binary systems, and 10 of the least massive are below, with all ordered by primary mass. In general, larger TNO binaries present a larger size disparity between the components and tend to be on tighter orbits (in units of the primary radius, R$_{\mathrm{primary}}$). Less massive binaries are more widely separated, with comparably sized components. This dichotomy implies two different formation mechanisms for the two observed populations of TNO binaries. For Pluto, only Charon is shown due to the very small sizes of the minor satellites. System masses, semi-major axes, and component diameters were retrieved from the Johnston Archive (\url{http://www.johnstonsarchive.net/astro/astmoontable.html}) and references therein.\label{binaries}}
\end{center}
\end{figure}

Ten of the largest known TNOs with at least one known satellite all have small secondaries with respect to the primary and separations $<$100 primary radii. Conversely, ten of the smallest known TNOs with known satellites have components of comparable size and a majority are separated by $>$100 primary radii (Figure~\ref{binaries}). This dichotomy suggests different formation mechanisms, of which three broad categories exist: capture, gravitational collapse, and giant impacts \citep[e.g.,][]{Brunini2020}. The capture mechanism has been well-studied \citep[e.g.,][]{Goldreich2002,Weidenschilling2002, Funato2004,Astakhov2005,Lee2007,Schlichting2008,Kom11}, but is not favored for the formation of small TNO binaries due to the preponderance of prograde orbits \citep{Grundy2011,Grundy2019}, which is less likely to result from capture \citep{Schlichting2008}, and the correlated colors of binary components \citep{Benecchi2009}. Instead, gravitational collapse by the streaming instability has gained traction due to its efficiency in creating binaries with equal-sized components and a wide range of semi-major axes \citep{Youdin2005, Johansen2009, Nesvorny2010, Simon2017, Li2018}. The giant impact mechanism is favored for the satellites of large TNOs \citep[e.g.,][]{Canup2005, Canup2011,Br06}, though capture \citep{Goldreich2002} and the collision of two objects within the Hill sphere of a third \citep{Weidenschilling2002} are also potential options.

The origin of the most massive TNO binary system, (136199) Eris and its satellite Dysnomia, is still under investigation. The giant impact and capture mechanisms are the more likely options, given that the \citet{Weidenschilling2002} mechanism mentioned previously is applicable primarily to the formation of small satellites \citep{Brunini2020}. The diameter of Dysnomia, \citep[700$\pm$115 km;][]{BB2018}, comparable diameter ratio of Eris/Dysnomia and Pluto/Charon (Figure~\ref{binaries}), and Dysnomia's low-eccentricity \citep[$e=0.0062\pm0.0010$;][]{H21}, prograde orbit all tend to favor a giant impact origin. However, the long orbital period of Dysnomia of 15.785899$\pm$0.000050 days \citep{H21}; stark albedo contrast between the two components, 0.96$^{+0.09}_{-0.04}$ for Eris \citep{Sicardy} vs 0.04$^{+0.02}_{-0.01}$ for Dysnomia \citep{BB2018}; and lack of information on Dysnomia's orbital inclination with respect to Eris' equatorial plane do not necessarily fit that paradigm.

In this work, we determine the rotation period of Eris in order to evaluate the tidal state of the system and understand its origins. Nearly two decades after its discovery, and despite being among the brightest TNOs, the literature still presents partial and conflicting determinations of Eris' rotation period. A wide range of rotation periods have been reported: $>$5 days \citep{Carraro2006}; 13.69, 27.38, and 32.13 hours \citep{Duffard2008}; 25.92 hours \citep{Roe2008}; and synchronous with Dysnomia's orbit\citep{RO2014,szakats}. \citet{Roe2008} identified a signal in their periodogram at $\sim$15 days, but discounted its significance because their photometry was obtained over a time period only twice as long.  The difficulty in determining Eris' period from its rotational light curve is that the amplitude of variation is very low, only $\sim$3\% (or 0.03~mag) peak-to-peak, and the period is long, such that accurate determination of the period requires high signal-to-noise ratio (SNR) and photometric calibration to accuracy better than 0.01~mag across months or years of observations. Furthermore, on this timescale the solar phase curve is important to consider when constructing periodograms from sparse data, even though Eris' solar phase varies from only 0.1 -- 0.6$^{\circ}$.

To meet this challenge, we combine photometry from three different telescopes. In chronological order: the Dark Energy Survey (\des) made 10 sweeps across a 5000~deg$^2$ swath of southern sky in each of the $g, r, i, z,$ and $Y$ filters over the period Aug 2013 through Feb 2019, using a large-format camera on the 4-meter Blanco telescope at Cerro Tololo Interamerican Observatory \citep{desdr2}. Usable images of Eris appeared in 8, 5, 6, and 7 exposures in the $griz$ bands, respectively ($Y$-band images have insufficient SNR to be useful), out of the $\sim$80,000 exposures of the \des\ Wide Survey. These exposures yield high SNR on Eris and benefit from the exquisitely accurate global photometric calibration of the survey \citep{fgcm}. These measurements are, however, too sparse in time to determine an unambiguous period for Eris on their own.

The second set of data are a collection of nearly 1000 60-second exposures obtained between Aug 2015 and Jan 2016 in the Johnson $V$ band with the ``Facility Optical Camera'' on the Palomar 60-inch (P60) robotic telescope \citep{Cenko2006}. Measures of useful SNR are obtained by averaging the exposures within each of the 72 nights. Their photometric quality is highly variable, but the \des\ imaging provides accurate reference magnitudes for objects in all the P60 exposures. The more rapid cadence over a shorter time span is complementary to the sparse, long-term cadence of the \des\ data in determining an accurate period, and the two overlap in time.

The third set of data are images from the the Hubble Space Telescope (\hst) with the \wfc/UVIS instrument through the {\it F606W} filter on 7 separate visits between 1 Jan 2018 and 3 Feb 2018 (GO program 15171, PI: B. Holler).  These yield very high SNR on Eris and easily resolve Dysnomia.  While the time span of the \hst\ data is too short to admit a precise determination of the photometric period, the high SNR and non-sidereal cadence of these data allow a more precise determination of the light curve and a veto on sidereal aliasing of the period.

As this work was being finalized, \citet{szakats} reported statistically significant candidate periods of $16.2\pm0.5$ days and $15.87\pm0.22$~days from a heterogeneous set of 31 nights of ground-based photometry spanning 15 years and \gaia\ DR3 $G$-band photometry spanning 2.5 years, respectively. Our results use a completely independent set of observations to test this assertion, and exploit higher precision observations and increased sampling to exclude possible aliases and obtain a measurement of Eris' rotation period that is $10\times$ more precise, as well as additional information on its phase curve and on Dysnomia's flux variations.  We derive Eris' rotation period using data independent of \citet{szakats}, and then incorporate the \gaia\ data into our final estimates of the light-curve parameters.

\section{Observations \& Data Reduction}
\subsection{Dark Energy Survey (\des) images}
The \des\ is a completed survey that covered $\approx 5000$ square degrees of the southern sky using the Dark Energy Camera (DECam) hosted at the Victor M. Blanco 4-meter Telescope at the Cerro Tololo Inter-American Observatory (CTIO) in Chile. A full description of the \des\ observing sequences and calibration steps is given by \citet{desdr2,despipeline} and \citet{fgcm}. We note that the relative photometric zeropoints of all the accepted exposures in the survey are determined very well by solving for consistency among the tight web of overlapping exposures taken during the survey. \citet{desdr2} demonstrate root-mean-square (RMS) differences of just 3~mmag between \des\ and \gaia\ stellar-source calibrations, implying that both surveys are calibrated to this level or better across the sky.

We extract photometry for Eris from the 26 relevant images in 2013--2018 using the methods for moving-object photometry described by Bernardinelli et al.\ (in preparation).
To summarize: photometric zeropoints and a model of the (color-dependent) point-spread function (PSF) for each exposure are derived as part of the survey pipelines. We model a small region around each exposure of Eris jointly with the (typically) 7 other \des\ exposures of that sky location in the same filter on different nights, when Eris was absent. The model consists of a free array of background sources that are assumed to exist in all exposures, plus a point source in Eris' location that is present only in a single exposure.  This ``scene-modeling photometry'' yields shot-noise-limited photometry for Eris that was unaffected by potential overlapping background sources. Following  \citet{desdr2}, we add an additional 0.003~mag of uncertainty to each measured Eris magnitude to allow for local zeropoint uncertainties. Resultant magnitudes for Eris are listed in Table~\ref{phottab}.

At maximum elongation, Dysnomia is $\sim$500 mas from Eris \citep{BS2007,BB2018,H21}, so the ground-based \des\ (and P60) images did not resolve Eris from Dysnomia. However, because Dysnomia's flux is 0.21$\pm$0.01\% that of Eris' at $\lambda\approx600$~nm (according to \citealt{BS2007}; we find higher but still small $\sim0.4\%$ values in Table~\ref{hstmag} ), any contamination of the \des\ or P60 light curves of Eris by variations in Dysnomia's magnitude must be at the sub-mmag level in the $g, r,$ and $V$ bands. In the $i$ and $z$ bands of \des\ the contribution of the redder Dysnomia could be much larger: \citet{Br06} reported Dysnomia to have 1.9$\pm$0.5\% of Eris' flux in the $K\arcmin$ band.

\begin{deluxetable}{cccccccc}[t]
  \tablewidth{0pt}
  \tablecaption{Photometric data}
  \tabletypesize{\small}
\tablehead{\colhead{MJD} & \colhead{Band} & \colhead{Phase ($^{\circ}$)} & \colhead{$\Delta$ (au)} & \colhead{r (au)} & \colhead{Mag} & \colhead{Noise (mmag)} & \colhead{SysError (mmag)}}
\startdata
56547.39494 & $g$ & 0.3441 & 95.63 & 96.45 & 19.0869 & 4.7 & 3.0 \\
56568.30048 & $g$ & 0.1823 & 95.49 & 96.45 & 19.0480 & 4.5 & 3.0 \\
56569.20107 & $g$ & 0.1765 & 95.49 & 96.45 & 19.0484 & 4.1 & 3.0 \\
56591.14610 & $g$ & 0.1704 & 95.49 & 96.44 & 19.0691 & 4.6 & 3.0 \\
56899.62065 & $G$ & 0.4492 & 95.69 & 96.39 & 18.6114 & 13.2 & \nodata \\
56899.69466 & $G$ & 0.4487 & 95.69 & 96.39 & 18.5931 & 13.9 & \nodata \\
56932.33489 & $r$ & 0.1910 & 95.43 & 96.38 & 18.4925 & 3.5 & 3.0 \\
\nodata & \nodata & \nodata & \nodata & \nodata & \nodata & \nodata & \nodata \\
58152.19177 & $F606W$ & 0.5497 & 96.49 & 96.15 & 18.8433 & 2.2 & 3.0 \\
58152.19766 & $F606W$ & 0.5497 & 96.49 & 96.15 & 18.8441 & 2.3 & 3.0 \\
58152.20492 & $F606W$ & 0.5497 & 96.49 & 96.15 & 18.8346 & 1.8 & 3.0
\enddata
\tablecomments{The photometric data used in determining Eris' light curve are given in temporal order. Magnitudes and MJDs are as observed, before corrections for distance and light-travel time. \texttt{SysError} is the amount added in quadrature to the measurement noise of each observation to account for calibration and other systematic errors. The \textit{F606W} band is from HST, $V$ band is from the Palomar 60-inch (with exposures already averaged into $\approx30$-minute segments), $g$ and $r$ are from \des, and $G$ band is from \gaia. Table 2 is published in its entirety in machine-readable format. A portion is shown here for guidance regarding its form and content.}
\label{phottab}
\end{deluxetable}

\subsection{Palomar 60-inch (P60) telescope}
The Palomar 60-inch (P60) telescope is a fully automated queue telescope that schedules observations in real time based on constraints for requested observations and sky conditions. In total, the Eris/Dysnomia system was observed by the P60 facility camera with the Johnson $V$ filter in 1054 exposures across 72 nights between 2015/08/06 and 2016/01/29. Average seeing at the P60 is 1.1$\arcsec$ in $R$ band in the summer and 1.6$\arcsec$ in the winter. The facility camera 
had a 2k $\times$ 2k back-illuminated CCD detector and a field of view of 12.9$\arcmin$ $\times$ 12.9$\arcmin$ \citep{Cenko2006}. The facility camera has since been replaced with the SED Machine \citep{SEDMachine}. Ultimately, 42 of the nights contained data that passed all quality cuts (i.e., adequate seeing, accurate pointing, photometric stability, etc.).

\begin{figure}[ht!]
\begin{center}
\includegraphics[scale=0.65,trim=0cm 0cm 0cm 0cm,clip=true]{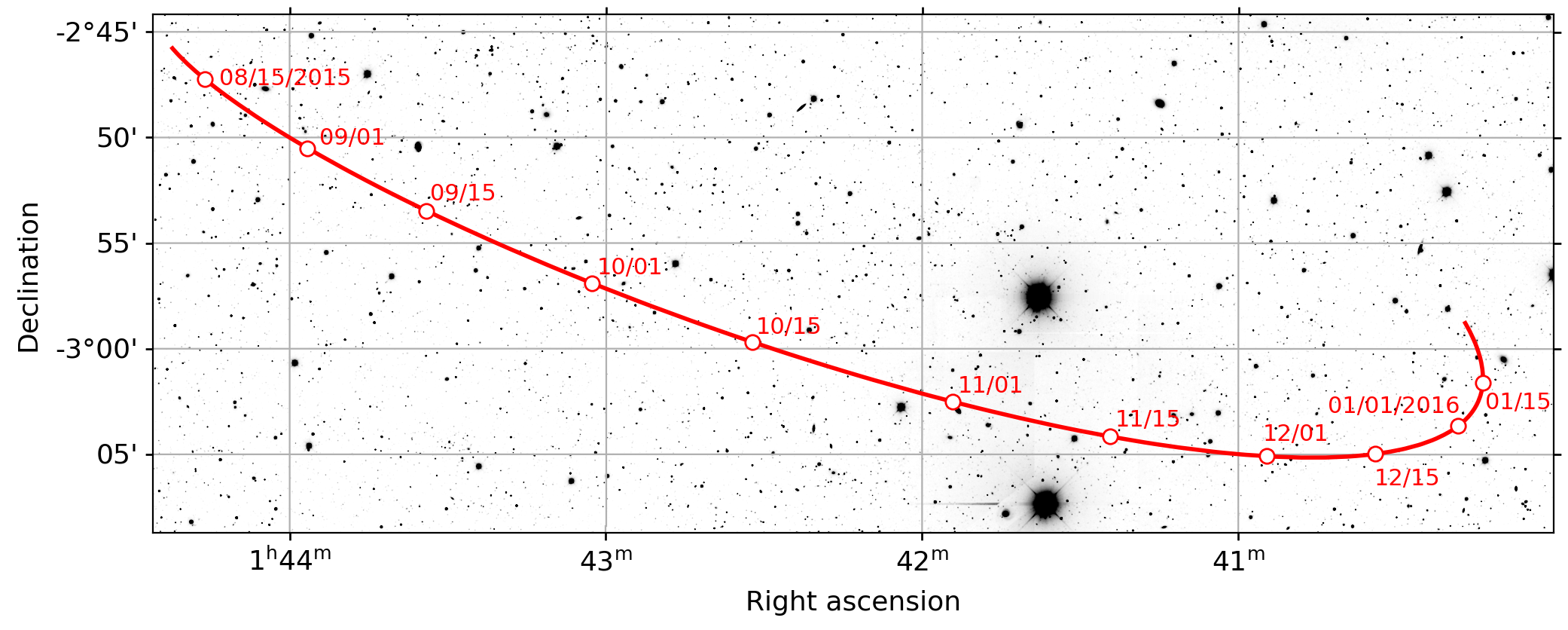}
\caption{Position of Eris from 2015-08-06 to 2016-01-29 UT, retrieved from JPL Horizons, with 15-day intervals marked. The background is the coadded \des\ $r$ band image with a scale of $\sim$1$\arcsec$/pixel. For comparison, the full field of view of the P60 facility camera is 12.9$\arcmin$ $\times$ 12.9$\arcmin$ \citep{Cenko2006}. \label{eris_pos}}
\end{center}
\end{figure}

A minimum of 12 1-minute images were requested each night. On some nights, additional sets of 12 images were obtained, but sometimes a sequence was partially or totally lost due to, e.g., a target-of-opportunity interruption or Eris being too near a bright star or detector defect. The track of Eris across the sky over this period is shown in Figure~\ref{eris_pos}. The track was wholly contained within the \des\ survey footprint, and we use the summed \des\ image (``coadd'') to confirm that a given sequence's images were not atop any other sources brighter than $r\approx24$~mag.

Raw images were immediately processed through the P60 image analysis pipeline, which handled demosaicking, overscan subtraction, bias subtraction, flat fielding with dome flats, sky-subtraction, bad-pixel masking, object detection, world coordinate system (WCS) construction, and seeing and photometric zeropoint estimation \citep{Cenko2006}. These processed data were then stored in an archive maintained by the Infrared Processing and Analysis Center (IPAC).

Extracting fluxes accurate to $<0.01$~mag from the P60 data requires several steps of processing and quality control. First, we determine instrumental  fluxes $f$ and uncertainties $\sigma$ for every star in every dome-flattened image via PSF fitting, as implemented by the codes \textsc{PSFEx} \citep{psfex} and \textsc{SExtractor} \citep{sextractor}. Measurements raising \textsc{SExtractor} error flags are discarded.

We identify detections of Eris by matching to its ephemeris, and the remaining P60 detections in each image are position-matched to stars in the \des\ coadd catalog (which is many times deeper than each P60 exposure).  The \des\ $g$ and $r$ magnitudes of each match are recorded. We discard any exposure which does not match at least 5 \des\ stars with $0<g-r<1.3$ and $S/N>10.$

The magnitude calibration process is done in batches of images from individual nights. We fit a zeropoint $m_{0,i}$ for each image $i$ and an overall color term $c$ for the night to the matched stellar images using the measured fluxes $f_{ij}$ and \des\ \texttt{PSF\_MAG\_APER\_8} magnitudes $g_j, r_j$ for star $j,$ via $\chi^2$ minimization to the model:
  \begin{equation}
    m_{0,i} - 2.5 \times \log_{10} f_{ij} = \tilde V_{DES,j} + c(g_j-r_j), \qquad \tilde V_{DES,j}\equiv (g_j+r_j)/2.
  \end{equation}
The synthetic $\tilde V$ band created from \des\ fluxes is close to the native Johnson $V$ band of the P60 data. We only use stars with $0<g_j-r_j<1.3$ for the photometric calibration.
During the fitting process we add 0.003~mag of estimated flat-fielding error in quadrature to the $\sigma_{ij}$ of individual stellar measurements, an amount chosen by eye to avoid over-weighting bright stars.  Then we iteratively clip measurements that are $>4\sigma$ away from the best fit.

We then fit the zeropoints for each night's exposures to a model of linear dependence on airmass $X$:
\begin{equation}
  m_{0,i} = m_0 + k (X_i-1),
\end{equation}
iteratively clipping individual exposures having residuals to this fit that exceed 3$\times$ the RMS variation of the night's residuals. We remove the clipped exposures from further consideration. We also drop the entire night's data if the RMS deviation from the airmass law exceeds 0.04~mag.
At this point we use the zeropoints $m_{0,i}$ and the color term $c$ to produce measured pseudo-$V$ magnitudes,
\begin{equation}
  \tilde V_{ij} = m_{0,i} - 2.5 * \log_{10} f_{ij} - c(g_j-r_j),
\end{equation}
for all surviving observations of Eris (taking $g-r=0.518$ for Eris from \des\ data) and the reference stars. We next split the night's exposures into segments of time spanning at most 40~minutes, and for each source we average all of the measurements taken in that time period into a single measurement.  We again perform sigma clipping to remove outliers caused by cosmic rays and other imaging defects. Most nights have only a single time segment. The output of the process is a catalog of time-averaged, calibrated $\tilde V$ magnitudes and their uncertainties for each source (including Eris), with both sky position and array position recorded, and the Julian date (JD) of the midpoint of each source's exposures.

At this point we must address a shortcoming of the P60 data reduction pipeline, which is the use of a dome flat to calibrate the response across the CCD. As emphasized in \citet{decamphot}, diffusely illuminated flat fields typically misrepresent the camera's response to stellar illumination because the flat fields count both focused and scattered light, whereas stellar photometry uses only the properly focused photons. The resultant photometric errors depend on position on the detector array. This is a serious issue for Eris' light curve because the pointing of the P60 images was held fixed for several months at a time, meaning that Eris moved across the array while the reference stars stayed fixed at one position.  Hence the flat-field errors are translated directly into a spurious variation in Eris' brightness.

To correct for flat-field errors, we create a ``star flat'' by first tabulating the residual errors $\tilde V_j-\tilde V_{DES,j}$ between P60 and \des\ magnitudes for each measurement of each useful reference star. These residuals populate the $(x,y)$ domain of the detector pixels irregularly. At each location on the detector, we set its star-flat value to the weighted mean of the 8 nearest reference-star residuals to that location, with the weights given by a Gaussian function of the distance. The star-flat image for the P60 data varies by 5\% across the detector, so these corrections are critical to obtaining useful light curve data for Eris. We assume that the star-flat correction is constant for the entire Eris campaign, so we use all valid exposures' stellar photometry to create it.

\begin{figure}
  \centering
  \includegraphics[width=0.9\textwidth]{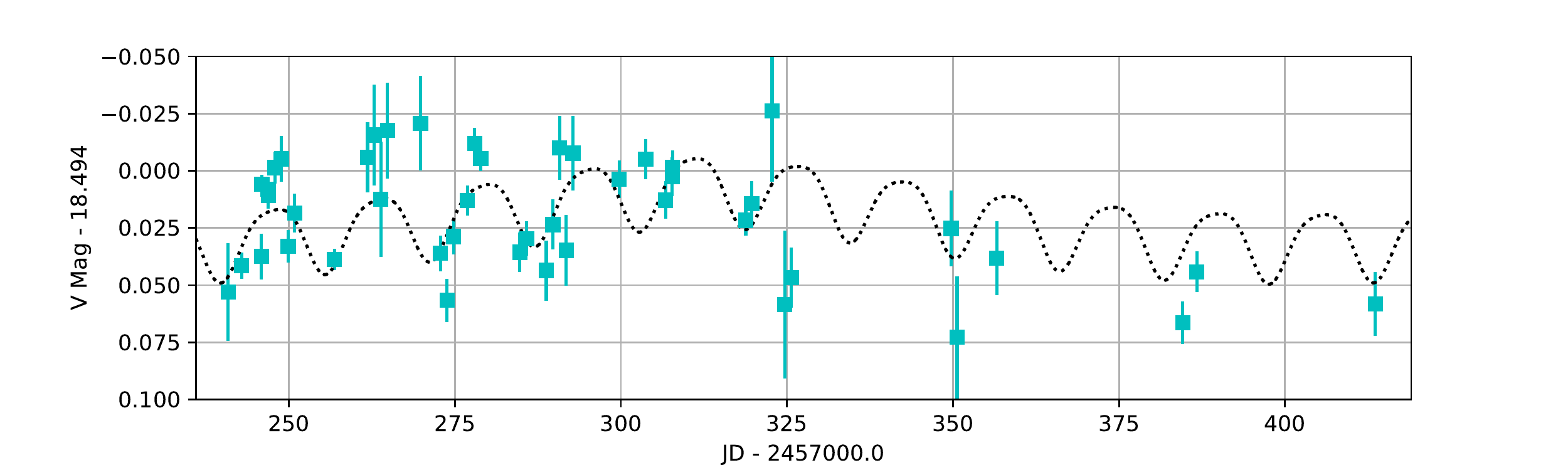}
\caption{Time series of the P60 $V$-band measurements. The error bars are statistical only, and do not include calibration errors or systematics. The dotted curve is the model for synchronous rotation and the solar phase curve derived in Section~\ref{fitsec}. The periodicity is marginally visible in these unbinned data---see Figure~\ref{phased} for phase-folded, binned versions.  It is also clear that the phase variations produce a signal of comparable amplitude to the rotational light curve.}
\label{p60time}
\end{figure}

The final step of the P60 photometry is to return to the original catalogs and adjust each stellar/Eris flux measurement by the value of the star flat at its detector location, and then repeat the entire calibration process. As a final quality-control step, we reject any measurement of Eris for which the $\chi^2$ per degree-of-freedom (DOF) for the individual exposures' magnitudes within a time segment is $>2.5.$ We drop the tilde and refer to these results as $V$ magnitudes henceforth. The final catalog (included in Table~\ref{phottab}) contains 45 measurements of Eris arising from $\approx450$ exposures on 42 distinct nights spanning 172 days, with a median uncertainty of 9~mmag. Figure~\ref{p60time} plots the results vs time.

\subsection{Hubble Space Telescope (HST)}
Each \hst\ visit was composed of four 348-second exposures and one 585-second exposure in a single orbit. Six visits were initially planned to occur within one Dysnomia orbital period, based on the 15.774-day period reported by \citet{BS2007}. However, visit 3 (2018/01/03) suffered a tracking failure so only two 348-second exposures were usable. An additional visit was awarded on 2018/02/03, at approximately the same orbital phase as visit 3, to offset these losses. The \wfc\ Instrument Handbook\footnote{\url{https://hst-docs.stsci.edu/wfc3ihb}} reports that the PSF full-width at half-maximum (FWHM) is $\sim$67 mas at 0.60 $\mu$m, resulting in just over 7 pixels between Eris and Dysnomia at maximum elongation. Additional details on these \hst\ observations can be found in \citet{H21}.
  
These images were retrieved from the Mikulski Archive for Space Telescopes (MAST) at the Space Telescope Science Institute (STScI). The raw images were reduced using the \wfc\ pipeline, \texttt{calwf3 v.3.6.2} (released 27 May, 2021)\footnote{\url{https://www.stsci.edu/files/live/sites/www/files/home/hst/instrumentation/wfc3/_documents/wfc3_dhb.pdf}}, which created a bad-pixel mask, corrected for bias, removed overscan regions, subtracted dark current, flat fielded, and normalized the fluxes between the separate UVIS1 and UVIS2 detectors. We make measurements on the \texttt{*flc.*} files which have had charge-transfer efficiency (CTE) corrections applied.

We fit each UVIS2 image to a model in which the signal $s_i$ in pixel $i$ at location $(x_i,y_i)$ is given by
\begin{equation}
  s_i = b + f_{\rm Dys}P(x_i-x_0-\Delta x, y_i-y_0-\Delta y,g) + f_{\rm Eris} \left[ P(x_i-x_0, y_i-y_0,g) \otimes D(R_{\rm Eris})\right].
\end{equation}
Here $P(x,y,g)$ is a PSF model for UVIS2 taken from tables created by \citet{BAG18}.\footnote{Because \hst\ was tracking Eris at a non-sidereal rate, it would have been inaccurate to use stars in the exposures as PSF models, even if there were enough to form a high-SNR PSF model.} We take the models for the {\it F606W} filter and interpolate them to the detector position of Eris's image. \citet{BAG18} find that the PSFs occupy a 1-dimensional family of shapes; they tabulate PSFs for 9 positions along this manifold. We assign a free parameter $g$ to the group number that applies to any particular exposure, allowing it to be a floating point value from 0 to 8, with the PSFs linearly interpolated between their groups. For Eris, we convolve the PSF with a circular disk of finite angular radius $R_{\rm Eris}$, with the brightness profile of a fully illuminated Lambertian hemisphere: $D \propto \sqrt{1-(r/R_{\rm Eris})^2}$. The known radius \citep{Sicardy} and distance of Eris imply a true angular diameter of 16.7~mas, or 0.42 \wfc\ pixels, but we leave $R_{\rm Eris}$ as a free parameter since we do not know the surface brightness distribution of Eris.  This could also be viewed as a generic adjustment of the PSF size to match the data. The background flux $b$, the fluxes $f_{\rm Dys}$ and $f_{\rm Eris}$, and the center in pixel coordinates $(x_0,y_0)$, of Eris are free parameters. The displacement $(\Delta x, \Delta y)$ from Eris' center to Dysnomia is taken from the orbit derived by \citet{H21}. We minimize the $\chi^2$ of the image data against the model with the free parameter set $\{f_{\rm Eris}, f_{\rm Dys}, b, x_0, y_0, g, R_{\rm Eris}\}$. The model is linear in the first three parameters, and we report uncertainties on the fluxes from these linear fits. Pixels affected by cosmic rays are excluded from the fits.

\begin{figure}
  \plotone{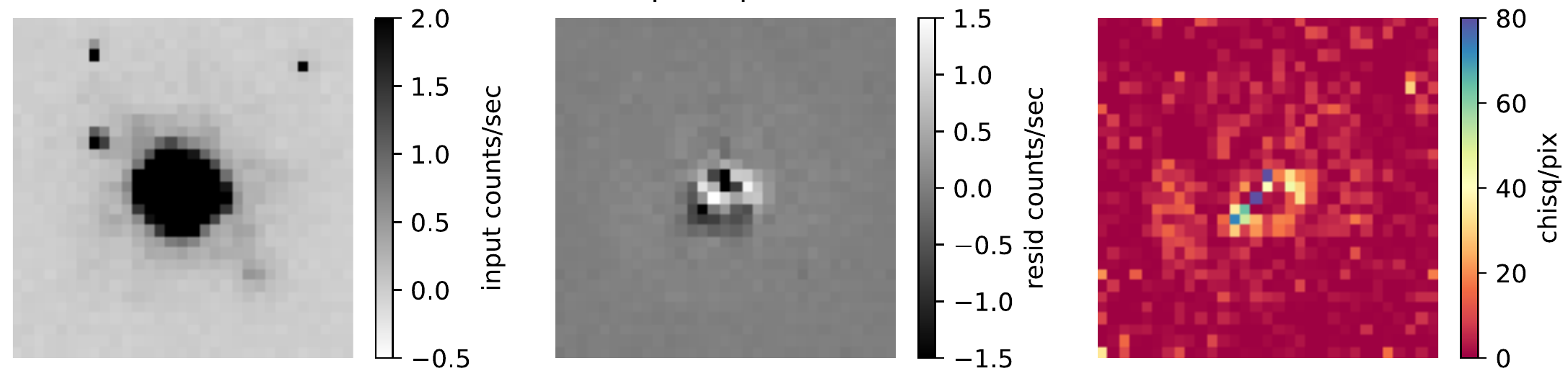}
  \caption{The model-fitting process for the \hst\ exposures is shown for a random exposure, \texttt{idl004icq}. At left is the processed image from MAST, with CTE correction applied. At center are the residuals to a model of a PSF for Dysnomia (to the lower right of Eris in this image) and a disk-broadened PSF for Eris. At right, the residuals are re-plotted as the $\chi^2$ per pixel using the pixel uncertainties reported by MAST. The residuals in the core of Eris are substantially larger than the statistical errors in the signal, due to the limitations of the PSF models.}
  \label{hstresids}
\end{figure}

As shown in Figure~\ref{hstresids}, the residuals to the model fits are well in excess of shot noise near the center of Eris, because the PSF models arre not sufficiently accurate. To reduce the errors in the derived flux that arises from PSF inaccuracies, we sum the residuals to the fit in the central $9\times9$ pixels of Eris' image and add them back into the model-fitting flux. In essence, we are using a simple aperture flux within this region and then using PSF fitting to infer the remainder of Eris' flux. 
The 32 resulting magnitudes for Eris are listed in Table~\ref{phottab}, along with their formal errors.

We expect the formal errors on Eris' flux (2.2~mmag for each of the shorter exposures) to be underestimates of the true uncertainty and derive an estimate of the additional systematic error in Section~\ref{fitsec}.

The $\chi^2$ minimizations yield $R_{\rm Eris}\approx0.70$~pix, significantly larger than the known physical value.  As noted above, this could be some combination of an inaccurate model for the disk's radial profile, or due to the PSF models being slightly too narrow, so we cannot draw any definitive conclusions about the nature of Eris' brightness distribution for this analysis.

The measurements of Dysnomia's magnitude have uncertainties of $\sim$0.05~mag for the shorter exposures. We find that in order for the exposures within each orbit to be mutually consistent with a magnitude that remains constant over the $\sim$45-minute duration of an orbit, we must add $\sim$0.04~mag of systematic error allowance in quadrature to the magnitude uncertainty derived for each exposure. Once this is done, and the exposures within each orbit are averaged to a single value, we have seven measurements of Dysnomia's light curve, with typical accuracy of 0.025~mag. These are presented in Table~\ref{hstmag}.

\begin{deluxetable}{cccccccc}[t]
\tablecaption{HST orbit-averaged photometry}
\tablewidth{0.5\textwidth}
\tablehead{\colhead{MJD} & \colhead{Phase ($^{\circ}$)} & \colhead{$\Delta$ (AU)} & \colhead{r (AU)} & \colhead{Eris (mag)} & Uncert. (mmag) & \colhead{Dysnomia (mag)} & Uncert. (mmag)}
\startdata
58119.28005 & 0.5725 & 95.95 & 96.15 & 18.5355 & 2.1 & 24.632 & 26 \\
58119.47863 & 0.5729 & 95.95 & 96.15 & 18.5389 & 2.0 & 24.589 & 25 \\
58121.25762 & 0.5764 & 95.98 & 96.15 & 18.5508 & 3.2 & 24.747 & 47 \\
58127.22729 & 0.5843 & 96.08 & 96.15 & 18.5437 & 2.0 & 24.466 & 25 \\
58128.41934 & 0.5847 & 96.10 & 96.15 & 18.5374 & 2.0 & 24.743 & 27 \\
58132.45865 & 0.5865 & 96.17 & 96.15 & 18.5330 & 2.0 & 24.486 & 24 \\
58152.19203 & 0.5497 & 96.49 & 96.15 & 18.5451 & 2.0 & 24.684 & 28
\enddata
\tablecomments{Magnitudes of Eris and Dysnomia measured by HST/WFC3/UVIS in the $F606W$ band. Each row is the combination of one HST orbit's exposures. Magnitudes and MJDs are as observed, before corrections for distance and light-travel time. Uncertainties include both measurement noise and estimated systematic contributions.}
\label{hstmag}
\end{deluxetable}

\subsection{Gaia}
Following \citet{szakats}, we extract observations of Eris from the \gaia\ DR3 \citep{gaiadr3,gaiadr3ss} table \texttt{gaiadr3\_sso\_observations}. There are 48 distinct observations spanning 2014--2017.  In order to maintain independence from the \citet{szakats} result, we do not use the \gaia\ data in fitting periods or initial light curves, but we do test whether the \gaia\ data are consistent with the light curve derived from the data sets discussed above. We do not attempt to assign a systematic error to \gaia's magnitudes. These data are included in Table~\ref{phottab}.

\section{Extraction of period and light curve for Eris}

\subsection{Method}
Given a set of measured magnitudes $m_i$ with uncertainties $\sigma_i$ taken at times $t_i$ in filter bands $b_i$, we search for periodicity at frequency $f$ by first adjusting the observation times for light-travel time and standardizing magnitudes to a (fictive) situation where the heliocentric and geocentric distances $r$ and $\Delta$ to Eris were at a reference distance of $d_0=90$~AU:
\begin{align}
  t_i & \rightarrow t_i - \Delta_i/c \\
  m_i & \rightarrow m_i - 5 \log_{10}(r_i \Delta_i / d_0^2).
\end{align}
We then fit the data to a model of sinusoidal variations including $N_h$ harmonics, and a linear dependence of magnitude on Eris' solar phase angle $\phi_i$ (measured in degrees) at each epoch:
\begin{equation}
  \hat m_i = m_{0, b_i} + \sum_{h=0}^{N_h} \left[ A_h \cos 2\pi ft_i  + B_h \sin 2\pi f t_i\right] +  G\phi_i.
  \label{model}
\end{equation}
This model is linear in the parameters $\{m_{0,b}\}$ for the mean magnitudes per band, the amplitudes $\{A_h, B_h\}$ per harmonic, and the phase slope $G.$  We scan across the nonlinear parameter $f$ to produce a periodogram of $\chi^2=\sum_i (m_i-\hat m_i)^2/\sigma_i^2$ vs frequency. Note that we assume that the light curve and phase slopes are the same in all of the bands. This is more likely to be true for the $g, r, G$, {\it F606W}, and $V$ bands, which are close to each other in central wavelength, than for the redder $i$ and $z$ bands. We examine the latter two bands after combining the former 5.

\subsection{Estimating the period}
\begin{figure}
  \begin{center}
  \includegraphics[width=0.7\linewidth]{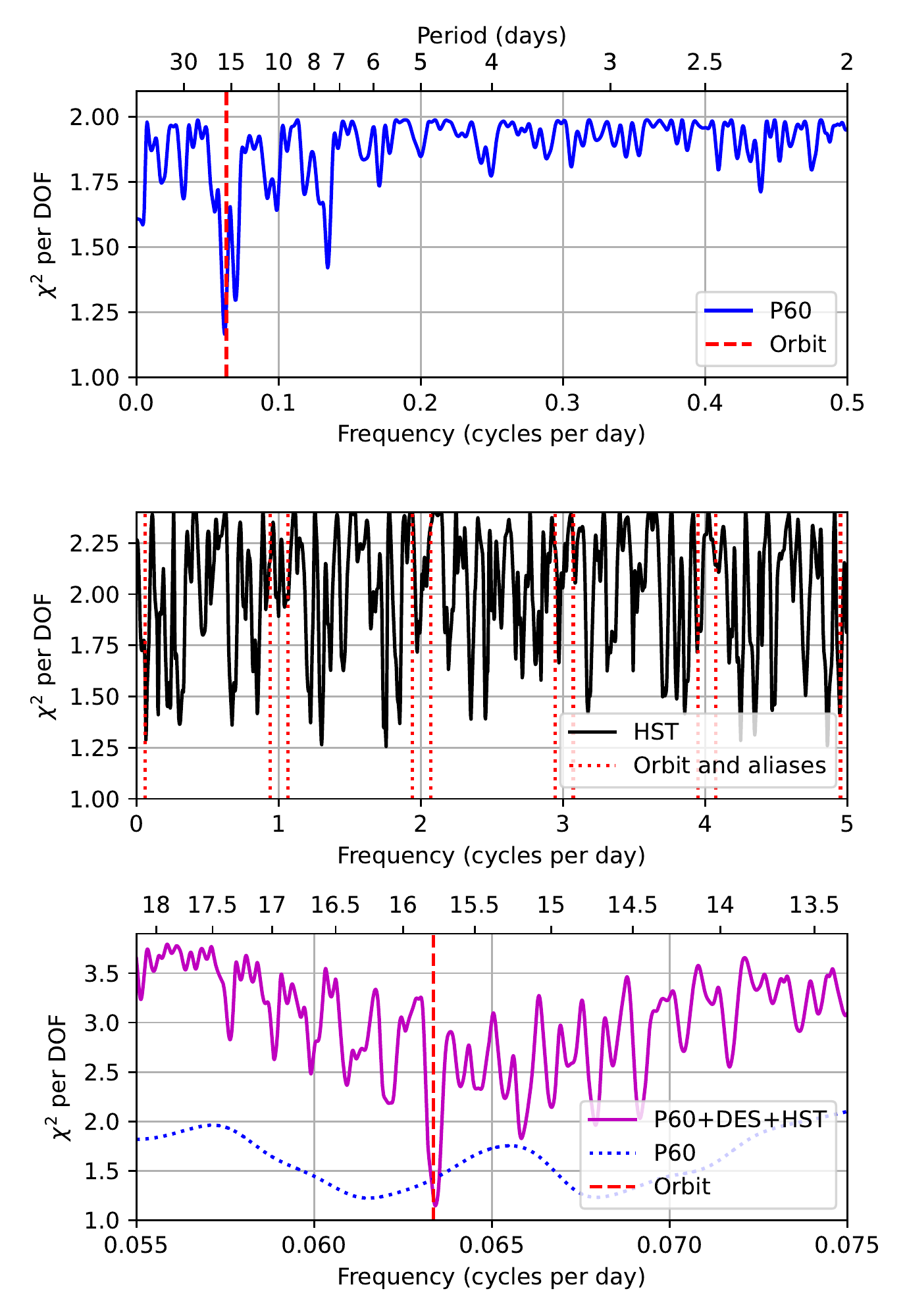}
  \end{center}
  \caption{Each panel shows the $\chi^2$ per DOF of a fit of a sinusoidal light curve to the photometric data vs frequency.  Ticks above the plots give period in days.
The upper panel uses the only the P60 data, revealing two strong minima, at periods of 16.1 and 14.4 days. The former is consistent with the orbital period of Dysnomia, $\sim$15.79~d, marked with the vertical dashed line.  Aliases of these frequencies due to the sampling of once per sidereal day are also good fits; we plot only the range below the Nyquist frequency.  The middle panel shows the $\chi^2/$DOF periodogram for the \hst\ data, which are not subject to sidereal-day aliasing.  The vertical lines mark the frequencies of synchronous rotation and all of its potential sidereal-day aliases with $P>5$~hr.  We can see that none of the possible $f>0.5/\textrm{day}$ aliases of the frequencies identified by the P60 data are good fits to a sinusoidal \hst\ light curve. The lower panel shows the periodogram from joint consideration of the P60, \hst\, and \des\ $g$ and $r$ band data, after estimated systematic errors are applied to each and the light curve is allowed to have a harmonic.  The data strongly select a photometric period consistent with Dysnomia's orbital frequency.}
  \label{pgrams}
\end{figure}

Because the P60 data are the only set that both resolved Eris' rotation period and spanned at least several cycles, we present their results in isolation first to identify plausible rotation periods. Fitting a simple sinusoid ($N_h=0$) yields the periodogram shown atop Figure~\ref{pgrams}. We display only periods of $0<f<0.5$~cycles/day because the data were taken at nearly the same sidereal time each evening (as were the \des\ data), hence signals at any other frequency would be strongly aliased into this range. Two strong minima corresponding to periods of 16.1 and 14.4 days are apparent. Including a harmonic, $N_h=1$, the same two peaks dominate, though additional minima appear (as expected) at half of the original two frequencies, and their aliases. The first period is fully consistent with synchronous rotation at the orbital period of Dysnomia. The best-fit values of the phase slope $G$ are in the range of 0.07 mag/degree, which is plausible.

The \hst\ data alone cover too short a time span to give a precise period, but they do offer multiple measurements at high SNR within roughly 2 cycles of the fundamental periods identified by the P60 data, and are not confined to fixed sidereal-time intervals. We use these in isolation to see if any of the aliases at $f>0.5$ cycles/day are better fits than the $\sim$15-day periods identified by the P60 data.  As shown in the middle panel of Figure~\ref{pgrams}, none of the aliases are consistent with the data, and we confine our further attention to the periods near 15 days indicated by the P60 light curve.

\subsection{Fitting combined data}
\label{fitsec}
The P60$+$\hst$+$\des\ data span $T\approx5$ years. If high-SNR data are available at either end of the time span, a shift in frequency of $\Delta f\approx 0.1/T$ will move the extremal points by 0.1~cycle on a phased light curve.  A shift of this size away from the true $f$ should therefore push at least one measurement to be a bad fit to the mean light curve.  We therefore expect the uncertainty in the derived period, $\Delta P$, to be roughly $\Delta P \approx \Delta f / f^2  = 0.1P^2/T\approx 0.02$~days.  To summarize, the data being fit included:
\begin{itemize}
\item 45 distinct observing segments in $V$ band from P60. We add in quadrature to each point's uncertainty an allowance of 12~mmag for systematic errors in photometry and calibration. This value is chosen to bring the $\chi^2$ per DOF near unity for the best-fitting light curves.
\item $8+5$ observations in $g$ and $r$, respectively, from \des. Each has 3~mmag of calibration systematic errors added in quadrature to its model-fitting error estimate, based on the comparison of \des\ stellar photometry to \gaia.
\item One mean magnitude for each of the 7 orbits of \hst\ data. Since the period is known to be much longer than the duration of a single HST orbit, we work with the per-orbit weighted-average magnitudes, which are listed in Table~\ref{hstmag}. We assign an independent systematic error of 3~mmag to each exposure before averaging; this is the value required to reduce the $\chi^2$ per DOF to near unity for a model in which the magnitude was constant during each orbit.
\end{itemize}

The model of Equation~\ref{model} is then fit to these 65 measurements.  Allowing for a single harmonic ($N_h=1$), there are 4 free parameters for the periodic terms, 4 free mean magnitudes in $g,r,V,$ and {\it F606W}; and one each for $f$ and $G.$ The lower panel of Figure~\ref{pgrams} shows the resultant periodogram, which now strongly prefers a period consistent with the system's orbital period.
If we formally estimate a $1\sigma$ confidence level as the range over which the $\chi^2$ value increases by unity over its minimum, we obtain $P=15.771\pm0.008$~days. The precision is in the range expected given the duration of the monitoring, and this $\pm2\sigma$ confidence region includes the orbital period of 15.785899$\pm$0.000050 days \citep{H21}.  The formal estimate of the uncertainty on the period is probably optimistic, given that we make some fairly crude allowances for systematic errors in the photometry, and assume a simple model for the light curve.  In any case we have strong evidence, independent of \citet{szakats}, that the rotation period of Eris is within 1 part in 1000 of Dysnomia's orbital period. Henceforth we will assume that Eris' rotation period is synchronous with Dysnomia's orbital period.

\begin{figure}
  \begin{center}
    \includegraphics[width=0.7\linewidth]{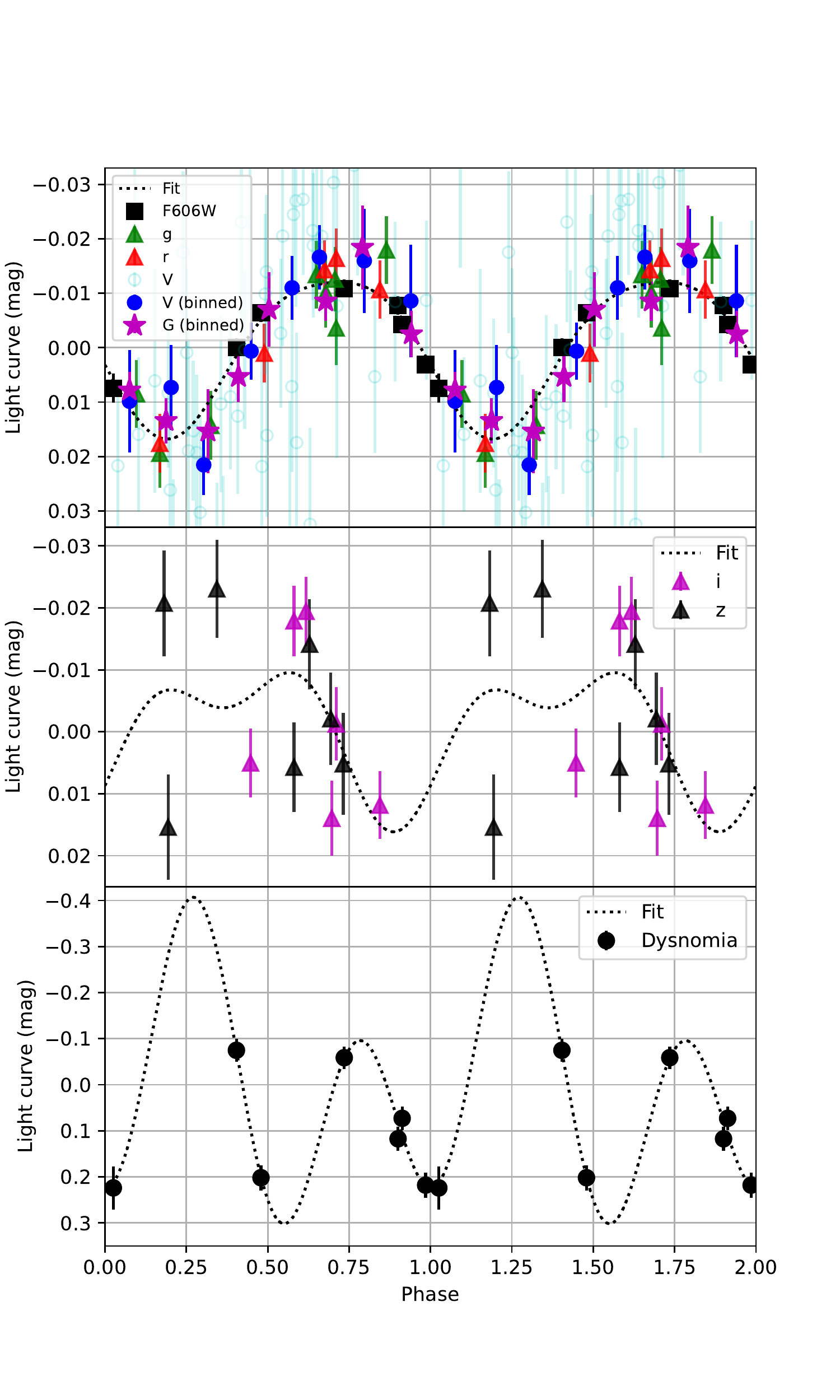}
  \end{center}
  \caption{All panels show light curves folded at the orbital period of Dysnomia. Mean magnitudes and illumination phase variation have been subtracted.  The top panel shows the \des\ $g$ and $r$ bands (triangles), HST {\it F606W} data (squares), and the individual P60 $V$-band data (light cyan circles). The dashed line is the best-fit 2-frequency light curve, with 0.03~mag peak-to-peak amplitude. The blue circles are the P60 data binned into 8 phase ranges. The magenta stars are the phase-binned \gaia\ $G$-band measures, which were \emph{not} used to derive this light-curve fit. The center panel shows the \des\ $i$ and $z$ bands---the dotted curve is the best-fit light curve, and the data are clearly not consistent with even this best periodic model; we suspect there are uncorrected systematic errors in the $iz$ photometry. The lower panel shows the HST magnitudes for Dysnomia. There is a clear variation of 0.3 mag, and they are plausibly fit by a double-peaked light curve from a fundamental and first-harmonic sinusoid, although there are only 7 points so this is certainly not conclusive.}
  \label{phased}
\end{figure}

The top panel of Figure~\ref{phased} shows the light curve, folded at the orbital period after removal of the mean magnitude and illumination-phase correction. The best-fit light curve has a center-to-peak amplitude of $0.150$~mag at the fundamental and only $0.002$~mag in the harmonic. Adding the harmonic does not induce a significant decrease in $\chi^2$, hence the light curve is very close to a single-peaked sinusoid. The best-fitting phase slope is $G=0.05\pm0.01$~mag/deg, (Table~\ref{fitparams}) indicating a stronger opposition surge than the slope of $G\approx0.035$~mag/degree measured for Pluto by \citet{Buie2010}, but weaker than the slope of $\approx 0.3$~mag/degree they observe for Charon at phase angles of 0.3--0.5$^{\circ}$.

The upper panel of Figure~\ref{phased} also includes the $G$-band \gaia\ fluxes (magenta stars), averaged into 8 phase bins. These agree perfectly in phase and amplitude with the light curve derived from the 3 other data sets, confirming the accuracy of the synchronous solution.

We remind the reader that the non-HST observations blend the flux of Dysnomia with that of Eris.  But near the $V$ band, Dysnomia's total light is only a $\approx2$~mmag perturbation to Eris's magnitude, so is an insignificant contributor to the light curve at current accuracy.

The \des\ $i$ and $z$ measurements are not well fit by a light curve with any sensible number of harmonics. The dashed curve in the center panel of Figure~\ref{phased} shows the phased data and the best-fit $N_h=1$ light curve. Indeed in each of these bands there are discrepant measurements at the same phase, suggesting measurement errors.  There may be an unknown source of systematic errors in these bands.  It is possible, for example, that Dysnomia's redder surface means that it is bright enough in $iz$ to confound the PSF-fitting results on Eris's flux in a seeing-dependent way.  We choose to ignore the $iz$ data since any inference from it would be questionable.

The lower panel of Figure~\ref{phased} shows the HST magnitudes for Dysnomia, averaged into HST orbits and phased at the orbital period. Similar to the HST Eris measurements, we determine the level of additional systematic error needed to make Dysnomia's fluxes within each orbit statistically consistent. This turns out to be about 0.04~mag, which we add in quadrature with the individual exposures' magnitude estimates before averaging by orbit. With only 7 measurements over 30~days, it is impossible to determine a photometric period. The {\it a priori} expectation is, however, that if Eris were in synchronous rotation then the less-massive Dysnomia would be as well. There is a clear detection of (at least) 0.3~mag peak-to-peak variability in Dysnomia's flux, and the light curve clearly is not sinusoidal at the orbital frequency.  The dashed line in this panel is a fit to the simplest double-peaked light curve (a fundamental plus first-harmonic sinusoid). This shows that a double-peaked light curve with a period equal to Dysnomia's orbital period is a plausible (though by no means unique) fit to the Dysnomia HST data.

Table~\ref{fitparams} gives the quantitative results of fitting models with one or two sinusoidal components to the light curve with all data.   Here it is apparent that the detection of the second harmonic is weak ($\Delta \chi^2\approx -7$ for 2 additional degrees of freedom).  The quantities of potential physical interest---period, light curve amplitude, and phase relation slope---are robust to the inclusion of the harmonic.  Because the assigned systematic errors on our photometry, and the light curve model, are approximations, the resulting uncertainties on fitted parameters may not represent an exactly 68\% confidence region.
The last row of Table~\ref{fitparams} shows the fitted parameters when the \gaia\ data are included in the fit.  All of the parameters shift by less than their estimated $1\sigma$ uncertainties.  For the purposes of this table, we change the parameterization from Eq.~\ref{model} to a form which isolates the total light curve semi-amplitude, $A$:
\begin{align}
  \hat m_i & = m_{0,b_i} + G\phi_i + A\cos(\theta_i) + C_2 \cos(2\theta_i) + S_2\sin(2 \theta_i) \nonumber \\
  \theta_i & = \frac{t_i - t_0}{P} \times 360^\circ - \theta_0.
             \label{newmodel}
\end{align}
The time $t_i$ is the time of emission of the light from Eris, and $t_0$ is the reference time JD 2457000.  The parameter $A$ gives the sinusoidal semi-amplitude of the fundamental, and $\theta_0$ is phase relative to $t_0$ at which the fundamental reaches its minimum.  $C_2$ and $S_2$ specify the harmonic signal, if included in the model.
\begin{deluxetable}{lccccccc}
  \tablewidth{0pt}
  \tablecaption{Fitted light-curve parameters}
\tablehead{\colhead{Model} & \colhead{Period (days)} & \colhead{$\chi^2 / \textrm{DOF}$} & \colhead{$G$ (mag/deg)} & \colhead{$A$ (mag)} & \colhead{$\theta_0$ (deg)} & \colhead{$C_2$ (mag)} & \colhead{$S_2$ (mag)}}
\startdata
With harmonic   &  $15.771 \pm 0.008$ & $68.4 / 56$ & $0.054\pm0.012$ & $0.015\pm0.001$ & $69\pm5$ &
$+0.003\pm0.001$ & $-0.001 \pm 0.001$ \\
Fundamental only &   $15.771 \pm 0.008$& $75.4 / 58$ & $0.054\pm0.012$ & $0.013\pm0.001$ & $68\pm4$ & $\equiv 0$ & $\equiv 0$ \\
Fundamental w/\gaia &   $15.769 \pm 0.007$& $131.4 / 105$ & $0.051\pm0.011$ & $0.013\pm0.001$ & $70\pm4$ & $\equiv 0$ & $\equiv 0$
\enddata
\tablecomments{Values and uncertainties of the parameters of a fit to the combined data of the form in Eq.\ref{newmodel}, with and without allowing for a second harmonic.  The first two lines omit \gaia\ data, thus matching the light curve shown in Figure~\ref{phased}, and are independent of \citet{szakats}; the third line adds \gaia\ to the fit to give more total constraining power. The uncertainties are given as the values that result in an increase of $\Delta\chi^2=1$ from the best-fit value, after normalizing such that $\chi^2/\textrm{DOF}=1$ at the best fit.  The uncertainties given are marginalized over all other free parameters.  For parameters other than $P$, the values and uncertainties are given with the period fixed to Dysnomia's orbital period.}
\label{fitparams}
\end{deluxetable}

\newpage 
\section{Discussion}

\subsection{Origin of the Eris-Dysnomia system}
The initial orbital and physical conditions of the Eris-Dysnomia system are dependent on the formation mechanism. A giant impact should create a Dysnomia interior to its current orbit, with subsequent evolution outward through the effects of tides. Conversely, if Dysnomia was captured, it is more likely that it started at a more distant semi-major axis and migrated inwards. If at any point Dysnomia's orbital period is shorter than (or retrograde to) Eris' rotation period, the tides raised on Eris by Dysnomia lag behind Dysnomia's position in its orbit, exerting a torque that transfers energy from Dysnomia's orbit to Eris and decreasing both Eris' rotation period and Dysnomia's semi-major axis. A more likely scenario is the opposite case, where Dysnomia's orbital period is longer than Eris' rotation period, transferring energy instead from Eris to Dysnomia, which increases both Eris' rotation period and Dysnomia's orbital period.  We quantify each of these scenarios using simplified equations for tidal evolution given by \citet{Goldreich68}:
\begin{align}
\label{omegaE}
\dot{\omega}_E & =\mathrm{sign}(\Omega-\omega_E)\frac{15}{4}\frac{k_E}{Q_E}\frac{m_D}{m_E}\left(\frac{R_E}{a}\right)^3\frac{Gm_D}{a^3} \\
\dot{\omega}_D& =\mathrm{sign}(\Omega-\omega_D)\frac{15}{4}\frac{k_D}{Q_D}\frac{m_E}{m_D}\left(\frac{R_D}{a}\right)^3\frac{Gm_E}{a^3}
\label{omegaD}
\end{align}
In the above equations, $\dot{\omega}_E$ and $\dot{\omega}_D$ are the change in the rotational spin frequencies of Eris and Dysnomia, respectively; $\Omega$ is the mean motion of Dysnomia ($2\pi/P$); $k_E$ and $k_D$ are the respective tidal Love numbers for Eris and Dysnomia; $Q_E$  and $Q_D$ are their tidal quality factors; $m_E$ and $m_D$ are their masses; $R_E$ and $R_D$ are their radii; and $a$ is the semi-major axis of Dysnomia's orbit.\footnote{The tidal Love number, $k$, is a dimensionless parameter that defines the rigidity of a body, i.e., how strongly the body will deform due to tidal forces. The tidal quality factor, $Q$, is a dimensionless measure of the deformation of an object divided by the energy dissipated via heat due to the deformation; high $Q$ values indicate a less efficient dissipation of energy due to tidal stresses. The ratio of the tidal Love number to the tidal quality factor gives the dimensionless rate of the internal energy dissipation for the body being considered, with smaller numbers indicating lower dissipation and slower orbital migration.}
A critical parameter will be the mass ratio $q=m_D/m_E$ of the system.
The system mass ($M_{\rm tot}=m_D+m_E=1.6466 \times 10^{22}$ kg), system density (2.43 g cm$^{-3}$), and current semi-major axis ($a_o=37,273$~km) were all taken from \citet{H21}. The Eris radius ($R_E=1163$~km) was taken from \citet{Sicardy}.
Other parameters are less precisely known.  \citet{BB2018} estimate a
Dysnomia radius $R_D=350\pm57.5$~km from their weak mm-wave detection of Dysnomia.
Nominal values for the tidal quality factor were taken as $Q_E=Q_D=100,$ and the tidal Love numbers for Eris and Dysnomia were calculated as described in \citet{MD2000}, with the rigidity of an icy body, $\mu$, taken to be 4 $\times$ 10$^9$ N m$^{-2}$ \citep{Hastings16}.  A nominal density for Dysnomia was assumed to be 1.2 g cm$^{-3}$, less than half of the system density. In this case, the nominal radius from \citet{BB2018} yields a mass ratio $q=0.014.$

These equations can be coupled with the conservation of the system's angular momentum to solve for the time evolution of the system given the (unknown) initial and (known) final states. This can be done numerically, following \citet{Hastings16}, as elaborated for the Eris-Dysnomia system in \citet{szakats}.  An analytic solution is available as well \citep[e.g.,][]{MD2000}, as we recapitulate in Appendix~\ref{analytic}, and yields the expected time interval $t_0-t_b$ between the birth of the system and the attainment of synchronous Eris rotation as
  \begin{align}
    t_0-t_b & \approx q^{-1} P_0 \left(\frac{a_0}{R_E}\right)^{5} \frac{1}{39\pi}\frac{Q_E}{k_E} \left| 1 - \left(\frac{a_b}{a_0}\right)^{13/2}\right|\\
            & = (12.5\,\textrm{Myr}) \times q^{-1} \left(\frac{Q_p}{100}\right) \left(\frac{0.1}{k_E}\right) \left| 1 - \left(\frac{a_b}{a_0}\right)^{13/2}\right|.
              \label{dtanalytic}
  \end{align}

\subsubsection{Outward migration}
\begin{figure}[ht!]
\begin{center}
\includegraphics[scale=0.55,trim=0cm 0cm 0cm 0cm,clip=true]{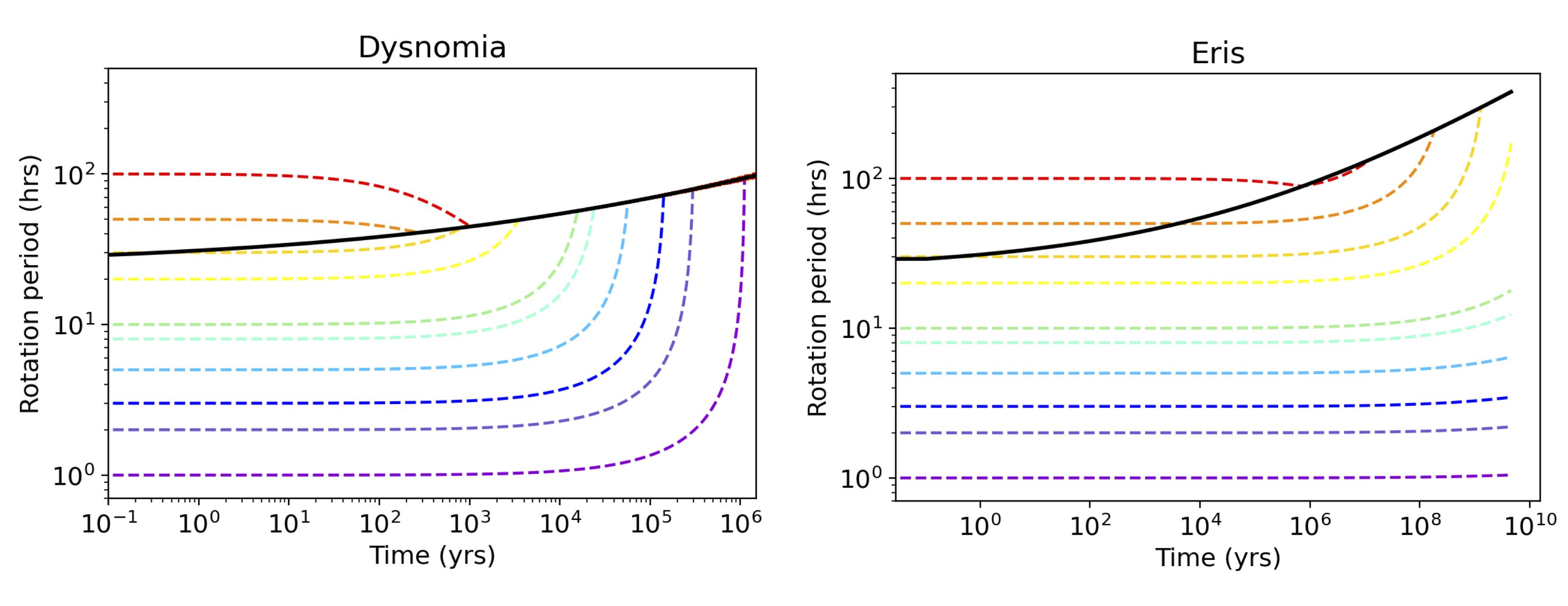}
\caption{Numerical models of the evolution of Dysnomia's (left) and Eris' (right) rotation periods, based on the method described in \citep{Hastings16}. Different initial rotation periods are represented by colored dashed lines and correspond to 1, 2, 3, 5, 8, 10, 20, 30, 50, and 100 hours. The black solid curve represents the synchronous period as a function of time in each plot.
Dysnomia's rotation and orbital periods synchronize in $\sim$1 Myr and Eris' rotation synchronizes for minimum initial periods comparable to that of the average singleton TNO ($\sim$10 hours). \label{eris_dys_num_model}}
\end{center}
\end{figure}

Figure~\ref{eris_dys_num_model} plots the history of $\Omega, \omega_D,$ and $\omega_E$ in numerical integrations of scenarios in which Dysnomia is formed interior to its current orbit.  Here we assume an initial semi-major axis of $a_b=6000$~km, $\sim$3$\times$ the nominal Roche limit. We consider a range of initial rotation periods from 1 to 100~hours for Eris.
The models indicate that Dysnomia's rotation and orbital periods synchronize in $\sim$1 Myr; this value varied less than an order of magnitude when considering densities ranging from 0.8 to 2.43 g cm$^{-3}$. For Eris, the rotation period synchronizes for initial periods comparable to that of the average singleton TNO (on the order of 10 hours; \citealt{LacLuu06,Thirouin14}). The minimum initial Eris rotation period that results in synchronization is highly dependent on the assumed density of Dysnomia. The nominal models assume a low density for Dysnomia of 1.2 g cm$^{-3}$, and the initial periods that result in synchronization vary from 2 hours to 30 hours for Dysnomia densities of 2.43 g cm$^{-3}$ and 0.8 g cm$^{-3}$, respectively. For the high-density Dysnomia case, the minimum initial period that results in synchronous rotation is shorter than Eris' critical rotation period\footnote{The {\it critical rotation period} is the rotation period that results in equivalent rotational and gravitational potential energies for a point mass on the surface of the body, and is defined as $P_{\rm crit}$ = $\sqrt{\frac{3\pi}{G\rho}}$. For $P<P_{crit}$, the rotational energy exceeds the gravitational potential energy and the body starts to break up. Note that this definition assumes a strengthless object; a real object with non-zero material strength would have a shorter critical rotation period, so this definition provides an upper limit to $P_{\rm crit}.$} of 2.12 hours, assuming Eris' density is comparable to the system density.

The numerical results are fully consistent with the expectations of the analytic solution in Eq.~(\ref{dtanalytic}). 
Requiring that synchronous rotation of Eris be attained in less than 4.6~Gyr implies a minimum mass ratio of Dysnomia to Eris of
\begin{equation}
  q > 0.003  \left(\frac{Q_E}{100}\right) \left(\frac{0.1}{k_E}\right).
\end{equation}
This bound is easily satisfied at the nominal $Q_E$ and $k_E$, since the measured 350~km radius
of Dysnomia implies a volume ratio of 0.027.  Thus Eris is easily brought into synchronous rotation in scenarios in which Dysnomia forms well interior to the current orbit, even with $Q$ as high as 1000 for Eris.

The hydrodynamic modeling performed to explain the formation of the Pluto-Charon system \citep{Canup2005}, and the strong resemblance of Eris-Dysnomia to Pluto-Charon, have implications for the formation of the satellites of other large TNOs (Figure \ref{binaries}) and the possible presence of other satellites in the Eris-Dysnomia system. The prevalence of other large TNO binary systems similar to Pluto-Charon and Eris-Dysnomia (e.g., Orcus-Vanth, Salacia-Actaea, and Varda-Ilmar{\"e}) implies a dynamically chaotic past for the trans-Neptunian region and a possible selection effect for the binaries observed today. It is possible that the only systems that survived were those that underwent a grazing collision, as proposed for the formation of the Pluto-Charon system, while objects that were more thoroughly collisionally disrupted would not have survived. The latter could have resulted in numerous collisional families scattered throughout the trans-Neptunian region, similar to those among the main belt asteroids.

The Pluto-Charon system is also home to four minor satellites in orbits beyond Charon \citep{Weaver06,Showalter11,Showalter12} that likely formed from the debris disk of the Pluto-Charon forming giant impact \citep{Canup2011}. A search for minor satellites of Eris using the same HST WFC3 data in this work was performed by \citet{Murray18}, with an imaging depth just capable of identifying satellites comparable to Nix and Hydra (the largest of Pluto's minor satellites) at the distance of Eris ($\sim$96 AU). No satellites were identified, meaning the formation of the Pluto-Charon system was the result of more unique circumstances, or any minor satellites around Eris are fainter than Nix and Hydra. Perhaps even more interesting is the claim that Eris may have an unresolved satellite interior to Dysnomia's orbit. This was proposed by \citet{Spencer21} to explain the statistically significant non-Keplerian nature of Dysnomia's orbit \citep{H21}, but can now be largely rejected given the rotation period determined in this work and in \citet{szakats}. Any satellite orbiting so close to Eris that it cannot be identified in HST WFC3 images, and that is massive enough to affect Dysnomia's orbit, would almost certainly have synchronized Eris to its own orbital period.

\subsubsection{Inward migration}
The radically different albedos and colors of Eris and Dysnomia might suggest a scenario in which Dysnomia was captured into a retrograde orbit at semimajor axis $a_b > a_0,$ and tidal migration was inwards.  In this case,
attainment of synchronous Eris rotation within 4.6~Gyr under Eq.~(\ref{dtanalytic}) implies
\begin{equation}
  \frac{a_b}{a_0} < 1.42 \left[ \left(\frac{q}{0.027}\right) \left(\frac{100}{Q_p}\right) \left(\frac{k_E}{0.1}\right) \right]^{2/13}.
\end{equation}
This $a_b$ is roughly $1\%$ of the Hill radius of Eris when it is at perihelion, 38~AU from the Sun.  We can also investigate the rotational period $P_{Eb}$ that Eris would need to have at the time of capture. Ignoring the spin angular momentum of Dysnomia as unimportant, conservation of angular momentum requires
\begin{align}
  1+\ell_p & = -\ell_p \frac{P_{0}}{P_{Eb}} + \left(\frac{a_b}{a_0}\right)^{1/2} \nonumber \\
  \Rightarrow \frac{P_{0}}{P_{Eb}} & = \ell_p^{-1} \left[ \left(\frac{a_b}{a_0}\right)^{1/2} -1 - \ell_p\right] \nonumber \\
           & \lesssim \ell_p^{-1}\left[\left(\frac {0.027\times 4.6\,\textrm{Gyr}}{12.5\,\textrm{Myr}}
             \right)^{1/13} - 1 \right] \nonumber \\
           & \approx 300 q \lesssim 10.
\label{pCapture}
\end{align}
In this evaluation we have set to unity the intrinsic physical quantity ratios that are raised to the $1/13$ power, and make use of the current ratio $\ell_p$ of Eris's spin angular momentum to the system orbital angular momentum (see Appendix~\ref{analytic}).
A pre-capture Eris rotation period of $P_{Eb}\approx10$~h, similar to the mean $<10$-hour periods reported for non-binary TNOs by \citet{LacLuu06} and \citet{Thirouin14}, would have $P_0/P_{Eb}=38.$  
According to Eq.~(\ref{pCapture}), such a large period change is only possible during
the modest $\approx40\%$ decrease in orbital radius that is achievable in $<4.6$~Gyr if we have $q>0.13.$ A Dysnomia mass this large is not consistent with the reported size ratio of Dysnomia to Eris.  The capture theory is difficult to reconcile with synchronous rotation of Eris unless some other event slowed Eris' rotation beforehand.

\subsection{The surface of Eris}

The measured peak-to-valley amplitude from the Eris light curve is small at $3\%,$ but non-zero. Stellar occultation timing on two chords from \citet{Sicardy} favor a spherical projected shape for Eris, but few-percent deviations from sphericity can probably not be excluded by these data.  Nonetheless, a non-spherical ellipsoidal shape for Eris would lead to a double-peaked light curve, which would mean that our detected photometric period would correspond to rotation at precisely half the orbital period of Dysnomia.  This seems highly unlikely given the nearly circular orbit of Dysnomia \citep[$e<0.01$;][]{H21}, so it seems more likely that Eris' variability is dominated by longitudinal variations in albedo.

Albedo variations on Eris' surface are highly plausible given the large variation of surface albedos on Pluto \citep{Buratti17}, and the high mean albedo of Eris \citep[$p=0.96^{+0.09}_{-0.04},$][]{Sicardy} suggests some form of frost generation, and seasonal nitrogen cycling is suggested by the models of \citet{Hofgartner2019}.

Dysnomia's orbit is currently inclined by $\approx45^\circ$ to the line of sight, so if this orbit is close to equatorial, Eris' rotation pole is sufficiently inclined to bring a substantial fraction of its surface in and out of view during rotation. The surface variation in albedo must clearly exceed 3\% to generate the light curve, but otherwise a wide range of dark-patch albedos, sizes, and geometries could be conjured to produce the observed light curve. The few-percent albedo difference between the regions of the surface that rotate into and out of view are small compared to the hemispherical differences observed on Pluto (e.g., \citealt{Buratti17}), suggesting that Eris lacks large-scale features as contrastive as Pluto's Sputnik Planitia and Cthulhu Macula.

\subsection{The surface of Dysnomia}

The double-peaked and high-amplitude Dysnomia light curve (lower panel of Figure~\ref{phased}) is what would be expected from a substantially asymmetric body in synchronous rotation, but Dysnomia's estimated diameter of $700\pm115$~km \citep{BB2018} is large enough that significant deviations from a sphere are not expected. Instead, this could indicate large dichotomies in Dysnomia's surface composition, manifesting as large dichotomies in color and albedo. Unfortunately, the HST data used to construct the light curve were obtained only through the $F606W$ filter, so no phase-dependent color information is available. Additionally, with the non-conclusive, sparsely sampled light curve and no occultation data to back up the assumption that Dysnomia is spherical, its surface properties remain entirely unconstrained, and it could in fact be ellipsoidal in shape with a double-peaked light curve. As a relatively bright TNO satellite around the most massive known TNO, Dysnomia is a prime target for the next level of study, which would include higher-cadence photometric observations and phase-resolved color and spectroscopic observations in order to understand the interactions between the two bodies and the ongoing evolution of the system.

\section{Summary}
We use three Earth-based data sets from the Palomar 60-inch telescope (high cadence), \des\ (long time baseline), and the Hubble Space Telescope (high photometric precision and non-sidereal sampling) to confirm the synchronous rotation of the TNO dwarf planet Eris first reported in \citet{szakats}. Highlights of the results and interpretations include:
\begin{itemize}
\item The rotation period of Eris is determined to be 15.771$\pm$0.008 days, within 2-$\sigma$ of the orbital period of Dysnomia determined by \citep{H21}, 15.785899$\pm$0.000050 days.
\item The amplitude of Eris' light curve is only 0.03 mag, suggesting that any large-scale albedo features such as Pluto's Sputnik Planitia and Cthulhu Macula that rotate in and out of the field of view have albedo variations of $\lesssim10\%.$  Smaller features with larger albedo variation are also plausible.
\item Eris' illumination phase slope of 0.05 mag per degree is between Pluto's and Charon's, implying a surface texture intermediate between those two objects.
\item The light curve of Dysnomia from HST WFC3 data is consistent with a synchronous period as well, but the small number of data points prevents a definitive determination. The large light curve amplitude of 0.3 mag (10$\times$ larger than Eris' amplitude) is consistent with an ellipsoidal object or a spherical object with large-scale surface dichotomies.
\item The formation of the Eris-Dysnomia system is best explained by a giant impact origin, including reasonable estimates for the initial rotation period of Eris, the initial semi-major axis of Dysnomia, and the density of Dysnomia. Tidal evolution of Eris due to the inward migration of Dysnomia after capture from heliocentric orbit is difficult to accomodate within the age of the solar system.
\end{itemize}

\section*{Acknowledgements}
The authors would first like to thank Richard Walters, Associate Research Engineer at the Palomar Observatory, for his help and patience in scheduling the imaging observations at the 60-inch telescope. The authors appreciate the work of Crystal Mannfolk, Linda Dressel, and Kailash Sahu of STScI in helping to optimize the HST observations prior to execution. Darin Ragozzine, Anne Verbiscer, Leslie Young, Michael Mommert, James Bauer, and Susan Benecchi provided helpful advice throughout this investigation. This work is based on observations made with the NASA/ESA Hubble Space Telescope, obtained from the data archive at the Space Telescope Science Institute. STScI is operated by the Association of Universities for Research in Astronomy, Inc., under NASA contract NAS 5-26555.

Support for this work was provided by NASA through grant number GO-15171.001 from STScI.  Work by GMB, PHB, and RNE was supported by National Science Foundation grants AST-2009210 and AST-2205808. PHB acknowledges support from the DIRAC Institute in the Department of Astronomy at the University of Washington. The DIRAC Institute is supported through generous gifts from the Charles and Lisa Simonyi Fund for Arts and Sciences, and the Washington Research Foundation.

Funding for the DES Projects has been provided by the U.S. Department of Energy, the U.S. National Science Foundation, the Ministry of Science and Education of Spain, 
the Science and Technology Facilities Council of the United Kingdom, the Higher Education Funding Council for England, the National Center for Supercomputing 
Applications at the University of Illinois at Urbana-Champaign, the Kavli Institute of Cosmological Physics at the University of Chicago, 
the Center for Cosmology and Astro-Particle Physics at the Ohio State University,
the Mitchell Institute for Fundamental Physics and Astronomy at Texas A\&M University, Financiadora de Estudos e Projetos, 
Funda{\c c}{\~a}o Carlos Chagas Filho de Amparo {\`a} Pesquisa do Estado do Rio de Janeiro, Conselho Nacional de Desenvolvimento Cient{\'i}fico e Tecnol{\'o}gico and 
the Minist{\'e}rio da Ci{\^e}ncia, Tecnologia e Inova{\c c}{\~a}o, the Deutsche Forschungsgemeinschaft and the Collaborating Institutions in the Dark Energy Survey. 

The Collaborating Institutions are Argonne National Laboratory, the University of California at Santa Cruz, the University of Cambridge, Centro de Investigaciones Energ{\'e}ticas, 
Medioambientales y Tecnol{\'o}gicas-Madrid, the University of Chicago, University College London, the DES-Brazil Consortium, the University of Edinburgh, 
the Eidgen{\"o}ssische Technische Hochschule (ETH) Z{\"u}rich, 
Fermi National Accelerator Laboratory, the University of Illinois at Urbana-Champaign, the Institut de Ci{\`e}ncies de l'Espai (IEEC/CSIC), 
the Institut de F{\'i}sica d'Altes Energies, Lawrence Berkeley National Laboratory, the Ludwig-Maximilians Universit{\"a}t M{\"u}nchen and the associated Excellence Cluster Universe, 
the University of Michigan, NSF's NOIRLab, the University of Nottingham, The Ohio State University, the University of Pennsylvania, the University of Portsmouth, 
SLAC National Accelerator Laboratory, Stanford University, the University of Sussex, Texas A\&M University, and the OzDES Membership Consortium.

Based in part on observations at Cerro Tololo Inter-American Observatory at NSF's NOIRLab (NOIRLab Prop. ID 2012B-0001; PI: J. Frieman), which is managed by the Association of Universities for Research in Astronomy (AURA) under a cooperative agreement with the National Science Foundation.

The DES data management system is supported by the National Science Foundation under Grant Numbers AST-1138766 and AST-1536171.
The DES participants from Spanish institutions are partially supported by MICINN under grants ESP2017-89838, PGC2018-094773, PGC2018-102021, SEV-2016-0588, SEV-2016-0597, and MDM-2015-0509, some of which include ERDF funds from the European Union. IFAE is partially funded by the CERCA program of the Generalitat de Catalunya.
Research leading to these results has received funding from the European Research
Council under the European Union's Seventh Framework Program (FP7/2007-2013) including ERC grant agreements 240672, 291329, and 306478.
We  acknowledge support from the Brazilian Instituto Nacional de Ci\^encia
e Tecnologia (INCT) do e-Universo (CNPq grant 465376/2014-2).

This manuscript has been authored by Fermi Research Alliance, LLC under Contract No. DE-AC02-07CH11359 with the U.S. Department of Energy, Office of Science, Office of High Energy Physics, and has gone through internal reviews by the DES collaboration.

This work has made use of data from the European Space Agency (ESA) mission
{\it Gaia} (\url{https://www.cosmos.esa.int/gaia}), processed by the {\it Gaia}
Data Processing and Analysis Consortium (DPAC,
\url{https://www.cosmos.esa.int/web/gaia/dpac/consortium}). Funding for the DPAC
has been provided by national institutions, in particular the institutions
participating in the {\it Gaia} Multilateral Agreement.

\newpage
\bibliographystyle{aasjournal}
\bibliography{eris}

\appendix
\section{Analytic solution for tidal locking timescale}
\label{analytic}
We summarize here an analytic solution to the most basic equations for the time to reach synchronous rotation of a satellite ($s$) and planet ($p$) to their orbit frequency $\Omega = 2\pi/P.$ The result for the synchronization time in Eq.~(\ref{tildet}) agrees with that given, for example, by Eq.~(4.123) of \citet{MD2000}. We assume the two bodies have masses $M_p$ and $M_s$ such that $M_p+M_s=M_{\rm tot}$ and $q=M_s/M_p.$  We assume a circular orbit at semi-major axis $a$ such that $GM_{\rm tot}=\Omega^2 a^3.$  If the radii are $R_p$ and $R_s$ we also have the density relation $q = \rho_s R^3_s / \rho_p R^3_p.$  The sum of the planet's and satellite's spin angular momenta and the orbital angular momentum is conserved, giving us:
\begin{align}
  \frac{L_{\rm tot}}{M_p} & = L_p + L_s + L_{\rm orb}(\Omega) \nonumber \\
                          & = L_p + L_s + \frac{q}{1+q} M_{\rm tot} \left(GM_{\rm tot}\right)^{2/3} \Omega^{-1/3}
                            \label{Lcons} \\
  \Rightarrow \tau_p + \tau_s & = \frac{M_{\rm tot}}{3} \frac{q}{1+q} \left(GM_{\rm tot}\right)^{2/3} \Omega^{-4/3} \dot\Omega,
  \label{dLdt}
\end{align}
where $\tau_p$ is the torque on the planet due to tides raised on it by the satellite, and $\tau_s$ vice-versa.
Eq.~(6) of \citet{Goldreich68} approximates $\tau_p$ generated on the planet by the tides raised on it from the satellite's mass as
\begin{align}
  \tau_p & = -\frac{3k_p}{2Q_p} \frac{ GM_s^2 R_p^5}{a^6} s_p, \nonumber \\
         & = -\frac{3M_{\rm tot}}{2} \frac{k_p}{Q_p} \frac{q^2}{(1+q)^2} R_p^5 \left(GM_{\rm tot}\right)^{-1} \Omega^4 s_p
           \label{taup} \\
  s_p & \equiv \textrm{sign}(\omega_p-\Omega).\nonumber 
\end{align}           
We introduce a dimensionless parameter giving the ratio of the torques (omitting the sign factors)
\begin{align}
  T \equiv \frac{\tau_s}{\tau_p} & = \frac{k_s}{k_p} \frac{ Q_p M_p^2 R_s^5}{ Q_s M_s^2 R_p^5} \nonumber \\
                                 & \approx \frac{\mu_p \rho_s g_s R_s}{ \mu_s \rho_p g_p R_p} \frac{ Q_p M_p^2 R_s^5}{ Q_s M_s^2 R_p^5} \nonumber \\
                                 & = \frac{\mu_p}{\mu_s} \frac{Q_p}{Q_s} \left(\frac{\rho_s}{\rho_p}\right)^{-1/3} q^{1/3} \label{tratio}.
\end{align}
In the second line we have taken the standard formula for the tidal Love number $k$ \citep[Eq. 3 from][]{Goldreich68} and approximated
\begin{equation}
  k = \frac{3}{2\left(1+\frac{19\mu}{2\rho g R}\right)} \approx \frac{3\rho g R}{19\mu},
\end{equation}
where $\mu$ and $g$ are the rigidity and surface gravity of the body, respectively, and we have approximated that the $\mu$ term dominates the denominator.  Note that $T$ is entirely determined by the ratios of intrinsic physical characteristics of the planet and satellite, times $q^{1/3}.$

Combining Eqs.~(\ref{dLdt}), (\ref{taup}), and (\ref{tratio}), we obtain a differential equation in $\Omega$ and its solution for the time interval between an initial state $i$ and final state $f$ as
\begin{align}
  \dot\Omega & = \frac{-9k_p}{2Q_p}q \left(GM_{\rm tot}\right)^{-5/3} R_p^5 \left(s_p + Ts_s\right) \Omega^{16/3} \\
  \Rightarrow \quad t_f - t_i & = \tilde t \left[ \left(\frac{\Omega_f}{\Omega_0}\right)^{-13/3}
                                - \left(\frac{\Omega_i}{\Omega_0}\right)^{-13/3} \right]  \left(s_p + Ts_s\right)^{-1}\\
             & = \tilde t \left[ \left(\frac{a_f}{a_0}\right)^{13/2}
               - \left(\frac{a_i}{a_0}\right)^{13/2} \right] \left(s_p + Ts_s\right)^{-1} \label{tsolve} \\
  \tilde t & \equiv \frac{2 k_p}{39 Q_p} q^{-1} R_p^5 \left(GM_{\rm tot}\right)^{5/3} \Omega_0^{-13/3} \nonumber \\
             & =  q^{-1} P_0 \left(\frac{a_0}{R_p}\right)^{5} \frac{1}{39\pi}\frac{Q_p}{k_p}, \label{tildet}
\end{align}
with $P_0$ and $a_0$ being period and semi-major axis of the orbit at a reference time, respectively, which we will make the present day for the Eris-Dysnomia system.

Eq.~(\ref{tsolve}) applies for any interval over which $s_p$ and $s_s$ are constant.  Consider a scenario for which the system is born at some time $t_b$ into an orbit with radius $a_b < a_0;$ then both planet and satellite spin down until a time $t_s$ when the satellite synchronizes to the orbit; then the planet spins down until a time $t_0$ at which the system is doubly synchronous and attains the current orbital configuration.  The time interval $t_b\rightarrow t_s$ has $s_p=s_s=1,$ and then switches to $s_p=1, s_s=0$ for $t_s<t<t_0.$ The total time to planet synchronization becomes
\begin{equation}
  t_0-t_b = (t_0-t_s) + (t_s-t_b) = \tilde t \left\{ 1 - \left(\frac{a_b}{a_0}\right)^{13/2} +\frac{T}{1+T} \left[\left(\frac{a_s}{a_0}\right)^{13/2} -  \left(\frac{a_b}{a_0}\right)^{13/2} \right]\right\}.
  \label{ttot1}
\end{equation}
The $13/2$ power of $a/a_0$ appearing in these solutions implies that the time to establishing a doubly-locked system is going to be $\tilde t$ to within a factor 2, unless the orbit has expanded $<10\%$ since satellite synchronization.

Should we wish to investigate in more detail and solve for the factor multiplying $\tilde t,$ we can introduce two more dimensionless parameters: $f_s\equiv L_{sb}/L_{pb}$ is the ratio of the spin angular momenta at birth, and we expect $f_s\ll 1$ for a system with $q\approx 0.01$, like Eris-Dysnomia.  The condition that the satellite synchronizes first is $f_sT<1.$
We also introduce $\ell_p \approx q^{-1} \left(a/R_p\right)^2$ as the ratio $L_p/L_{\rm orb}$ after synchronization.  If we take $L_s/L_{\rm orb}$ to be negligible, then the angular momentum conservation between the three epochs $t_b, t_s,$ and $t_0$ becomes
  \begin{equation}
    L_{pb}(1+f_s) + L_{\rm orb}(a_b) = L_{pb}(1-f_sT) + L_{\rm orb}(a_s) = (1+\ell_p) L_{\rm orb}(a_0)
  \end{equation}
  which, using $L_{\rm orb} \propto \sqrt{a},$ can be solved to yield
  \begin{equation}
    \left(\frac{a_s}{a_0}\right)^{13/2} = (1+f_s)^{-13} \left[ (1+\ell_p) f_s \frac{1-T}{1+T} + (1-f_sT)\sqrt{\frac{a_b}{a_0}}\right]^{13}.
    \label{asynch}
  \end{equation}
As long as $f_s<1$ and $a_b<0.8$ or so, the bracketed quantity in Eq.~(\ref{ttot1}) will be close to unity and the synchronization time will be $\tilde t.$


\allauthors

\end{document}